\documentclass[review,number,sort&compress]{elsarticle}

\pdfoutput=1

\usepackage{graphicx}
\usepackage{epstopdf}
\usepackage{subfigure}
\usepackage{amsmath}
\usepackage{amssymb}
\usepackage{afterpage}
\usepackage{graphics}
\usepackage{float}
\usepackage{rotating}
\usepackage{xspace}
\usepackage{dcolumn}
\usepackage{bm} 
\usepackage{lineno}
\usepackage{placeins}
\usepackage{changepage}
\usepackage{notoccite}

\journal{Nucl.Inst.Meth.A}

\begin{document}


\begin{frontmatter}



\title{Differences in the Response of Two Light Guide Technologies and Two Readout Technologies After an Exchange of Liquid Argon in the Dewar}
\tnotetext[t1]{FERMILAB-PUB-19-624-E-ND, arXiv: 1912.05987}

\author[label1]{S.~Mufson}
\address[label1]{Indiana University, Bloomington, Indiana 47405, USA}

\author[label1]{B.~Adams}

\author[label1]{B.~Baugh}

\author[label1]{B.~Howard\fnref{jose}}\fntext[jose]{currently: Fermi National Accelerator Laboratory, Batavia, Illinois 60510, USA}

\author[label1]{C.~Macias}

\author[label2]{G.~Cancelo}
\address[label2]{Fermi National Accelerator Laboratory, Batavia, Illinois 60510, USA}

\author[label2]{E.~Niner}

\author[label2]{D.~Totani}

\date{\today}          

\begin{abstract}
In this investigation the response to the scintillation light generated by through-going cosmic muons in liquid argon (LAr) was measured by two light guide technologies and two readout technologies after five weeks of running in the TallBo dewar at Fermilab.  The response was remeasured after the dewar was drained of LAr, refilled, and then run again for an additional four weeks.  After the dewar was refilled, there was clear evidence that the scintillation signal had dropped significntly.  The two light guide technologies were developed at Indiana University and MIT/Fermilab.  The two readout technologies were boards that passively or actively ganged 12 Hamamatsu MPPCs.  Two possible explanations were identified for the degraded signal: the response of the two light guide technologies degraded due to damage caused by thermal cycling, and/or unknown differences in the trace residual Xe contamination in the fills of LAr led to the observed rop in scintillation light.  Neither absorption nor quenching by N$_2$, O$_2$, and H$_2$O contamination can account for the degradation.  Neither the individual Hamamatsu MPPCs nor the passive/active ganging boards appear to have been affected by the thermal cycling.  The path length distributions of the cosmics traversing the dewar appear quite similar in both event samples.

\end{abstract}

\end{frontmatter}



\section{Introduction}
\label{sect:intro}

As charged particles traverse large liquid argon (LAr) detectors, the scintillation light they generate provides important information for investigations of neutrinos and dark matter.  To collect these scintillation photons for analysis, different technologies have been developed for the single phase far detector of the Deep Underground Neutrino Experiment (DUNE)~\cite{bib:DUNE-CDR-vol1}.  Two of these light guide technologies were studied in this experiment~\cite{bib:howard,bib:MITbars,bib:MITbars2}.  Both of these technologies make use of multiple silicon photomultipliers (SiPMs) in their readout.  Since the individual readout of multiple SiPMs is quite costly, two different ganging schemes were also studied in this experiment -- one in which the SiPMs were ganged by passively connecting the photodetectors in parallel and one in which the SiPMs were actively summed electronically.  

The questions addressed here ask what effects, if any, does thermal cycling have on this subset of technologies developed to detect the scintillation light in LAr.  Are these effects influenced by the conditions in the dewar or the contaminants in the LAr?  The experiment took place in the TallBo liquid argon dewar at Fermilab and studied single through-going, minimum ionizing cosmic muons.  Thermal cycling resulted when the dewar was drained after five weeks of running to remount the detectors, refilled, and then run again for an additional four weeks.  Since scintillation light can be significantly quenched by contaminants in the LAr~\cite{bib:MITN2,bib:O2Contamination,bib:O2contamCross,bib:H2O,bib:H2OcontamCross}, the contaminants were carefully monitored during the experiment.

\FloatBarrier
\section{The TallBo Experiment}
\label{sec:TallBoExperiment}

The experiment took place in the liquid argon dewar facility ``TallBo'', housed in the Proton Assembly Building (PAB) at Fermi National Accelerator Laboratory (FNAL).  The experiment ran from August 3, 2018 until October 17, 2018.  Data collection took place in 4 separate runs, which were defined by the detector configuration read out.  

\subsection{Experimental Design}
\label{sec:ExperimentalDesign}

\begin{figure}[h],
\centering
\includegraphics[width=3.8in,height=2.2in]{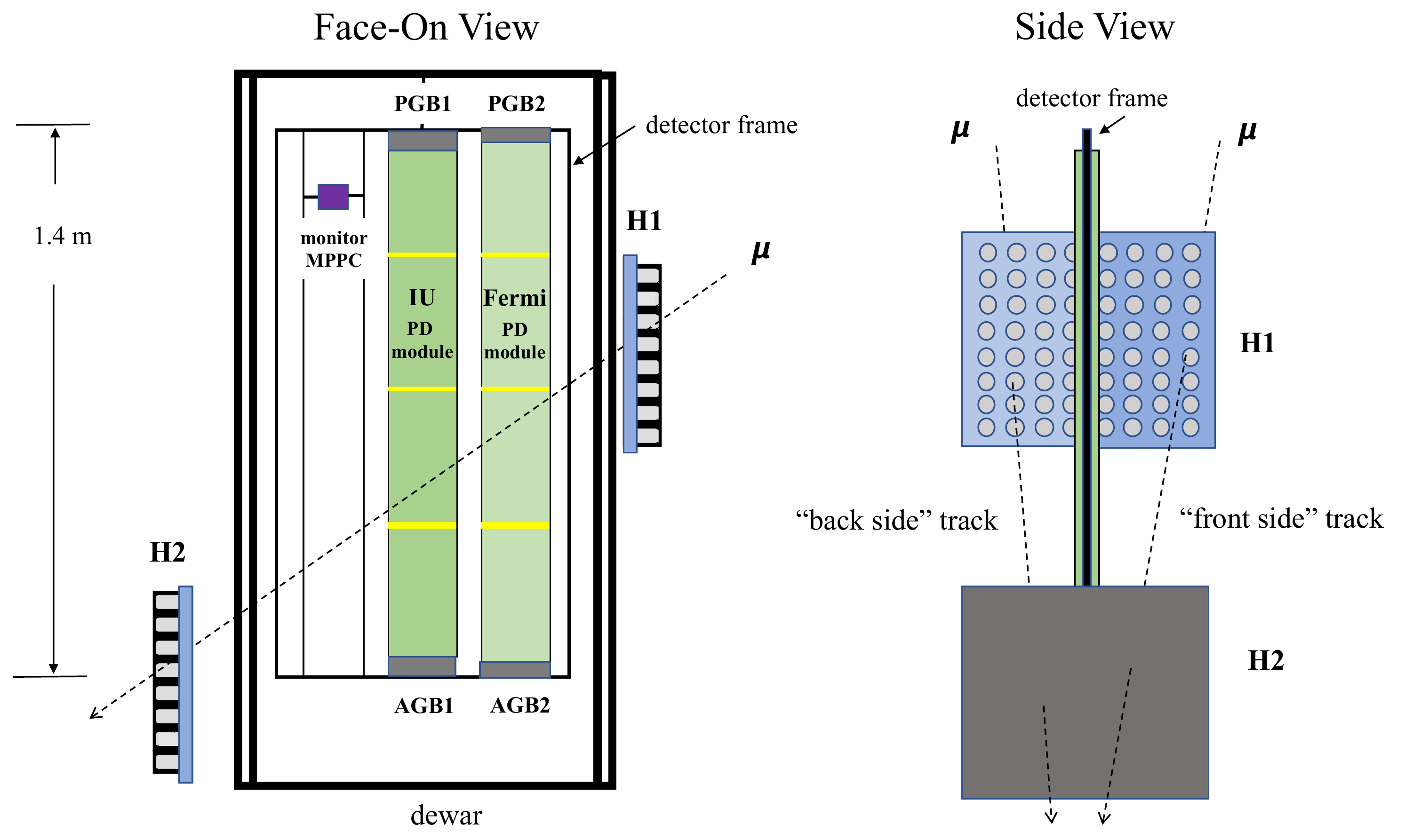}
\caption{Configuration of the experimental apparatus in the TallBo dewar during runs 1-3.  On the left is a face-on view with a representative single cosmic muon track that triggers the readout.  On the right is a side view, rotated by 90$^\circ$ counterclockwise, with representative ``front-side'' or ``back-side'' muon tracks that trigger the readout.  
}
\label{exptLayout}
\end{figure}

Fig.~\ref{exptLayout} shows two views of the experimental design in runs 1, 2, and 3.  

On the left is a face-on view of the experiment with a representative single cosmic muon track that would trigger the readout.  On the right is a side view, rotated by 90$^\circ$ counterclockwise, with two representative muon tracks that would trigger the readout.  These tracks are wholly on one side of the detector frame or the other.  Tracks that cross from one side of the detector frame to the other were excluded from the analysis.

The left of Fig.~\ref{exptLayout} shows two photon detector (PD) modules, one developed at Indiana University~\cite{bib:howard}, ``IU'', and one developed at MIT and implemented at Fermilab~\cite{bib:MITbars,bib:MITbars2}, ``Fermi''.  These detectors are mounted in a custom frame that is suspended from the dewar lid.  Each end of both PD modules was read out by 12 Hamamatsu SiPM (MPPC) photodetectors ganged together into one readout channel.  Two schemes were used to gang the MPPCs, passive and active.  The passive ganging boards, PGB1 and PGB2, read out the modules from the top of the frame and the active ganging boards, AGB1 and AGB2, read out the modules from the bottom.  Except for a few brief excursions (described below), the readout boards PGB1 and PGB2 were continuously submerged in LAr; the readout boards AGB1 and AGB2 were submerged in LAr uninterrupted during the entire experiment.  In run 4, the IU and Fermi modules were exchanged.  The readout boards, however, remained in the configuration shown in Fig.~\ref{exptLayout}.  

For an event to trigger the DAQ, coincidence signals were required from the two hodoscope modules, H1 and H2, that flank the outside of the dewar, as pictured in Fig.~\ref{exptLayout}.  The trigger signals from H1 and H2 were each composed of two components.  One component was generated from an 8$\times$8 array of 3$''$ PMTs, each of which was covered by a 2$''$ diameter barium-fluoride crystal coated with TPB.   The second component was generated by a scintillator panel placed between the PMT array and the dewar.  The scintillator panels on both H1 and H2 were comprised of two adjacent plastic scintillator sheets.  Each of the four scintillator sheets was read out by a single PMT.  The DAQ trigger required a 4-fold coincidence: signals over threshold from a PMT in H1 and in H2 , and signals over threshold from a scintillator panel in H1 and H2, all within a defined gate.  The thresholds and gates were set to reject large showers that would have overwhelmed the DAQ.

After installation, the readout PMT on one of the two scintillator sheets on H2 failed, while the other remained functional.  The PMT could not be replaced because spares were not available.  Consequently, tracks on only one side of H2 could generate a four-fold coincidence trigger.  As indicated in Fig.~\ref{exptLayout}, these tracks have been arbitrarily labeled  "front side" or "back side", depending on which side of H2 the functional scintillator sheet was mounted.  Front side tracks were recorded in runs~1, 2, and 4.  Back side tracks were recorded in run~3.

A single MPPC, mounted in the position shown, was included on the frame to monitor photodetector performance characteristics during the experiment.


%
%
In the IU light guide technology~\cite{bib:howard}, scintillation photons from liquid argon at 128~nm strike one of four acrylic plates with the wavelength shifter TPB embedded in their surfaces.  The TPB in the struck plate converts VUV scintillation photons to visible photons, typically in the range 420 -- 450~nm.  These  photons are transmitted through the plate and are subsequently absorbed by a commercial light guide made by Eljen Technology.  In the Eljen light guide, photons are again wavelength shifted to the range 480-510~nm and channeled to the Hamamatsu MPPCs at the end, where they are detected. 

The Fermi light guides~\cite{bib:MITbars,bib:MITbars2} were manufactured from cast acrylic bars that have TPB embedded into their surface using a dip-coating technology.  Scintillation photons from liquid argon are absorbed by the TPB and wavelength shifted to the range 420 -- 450~nm.  These waveshifted photons are channeled to the MPPCs on the ends of the light guides, where they are detected.

%
The SiPMs used in the experiment, including the monitor MPPC, were Hamamatsu 6x6mm MPPCs (S13360-6050VE\footnote[1]{https://www.hamamatsu.com/resources/pdf/ssd//s13360-2050ve$\_$etc$\_$kapd1053e.pdf}).
These MPPCs have 50$\mu$m pixels in a TSV package and are coated with epoxy resin.  
Fig.~\ref{testMPPCAppendix} in the Appendix shows the dark spectrum of a previously uncooled MPPC in LN2 that has been biased at 44.5~V.  The single photoelectron (p.e.) peaks and the signal from afterpulsing are marked. 
The MPPCs were biased at 44.5~V, a compromise that minimizes afterpulsing while maximizing the amplitude of the first p.e. peak.  This bias voltage is close to the 3~V over breakdown recommended by the Hamamatsu data sheets for operation at room temperature.  
The determination of the breadkdown voltage is described in the Appendix.
Laboratory measurements of the the performance characteristics of 12 previously uncooled Hamamatsu MPPCs immersed in LN2, biased at 44.5~V, are shown in Fig.~\ref{MPPCcharacteristicsAppendix} in the Appendix.  
There are 12 ganged MPPCs on each end of the light guides.  By ganging the MPPCs into one readout channel, the number of electronics channels and associated cables are significantly reduced, which in turn implies large reductions in electronics costs.  There were two different ganging board designs in the experiment -- passive and active.  
The signals from the readout boards were processed by a 12-channel SiPM Signal Processor (SSP) module that was designed and built by the HEP Electronics Group at Argonne National Laboratory (ANL).  The SSP has been described in~\cite{bib:TallBo}.

%
The passive ganging boards, PGB1 and PGB2 in Fig.~\ref{exptLayout}, connect the 12 MPPCs in parallel.  Parallel ganging increases the output capacitance proportional to the number of MPPCs connected in parallel but leaves the MPPC bias voltage unchanged.  Although the larger capacitance reduces the signal to noise ratio of the output, the integrated signal from the 12 MPPCs on the board remains constant.  The larger capactance also decreases the signal to noise on the waveforms during the increased signal collection time, particularly important for weak events.   Panel (a) in Fig.~\ref{boardWFsAppendix} shows the baseline subtracted mean of 25 waveforms from typical single track muon events crossing TallBo as read out by the passive board PGB1 during run 2.  The waveforms from the readout board PGB2 on the Fermilab technology are qualitatively very similar.  
%
%

One method to add the signals from the 12 MPPCs without increasing the capacitance as significantly is to use an active ganging board that implements an active summing node like the circuit shown in the {\it top} panel of Fig.~\ref{activeBoardDesignAppendix}.  
The active summing board that incorporaates the summing node is shown in the {\it bottom} panel of Fig.~\ref{activeBoardDesignAppendix}.
Panel (b) in Fig.~\ref{boardWFsAppendix} shows the mean of 25 waveforms from typical single track muon events crossing TallBo as read out by the active board AGB1 (Fig.~\ref{exptLayout}) during run 2.  The waveforms from the readout of the Fermilab technology are qualitatively very similar.  Clearly this board introduces a significant overshoot in the waveforms.  The overshoot is the most problematic artifact when analyzing their summed waveforms.  An artifact is also seen at the waveform minimum from the active ganging board.  Like the overshoot this artifact is the likely result of mismatches in impedance at the board-SSP coupling.

\FloatBarrier
\section{Operations}
\label{sec:operations}

To prepare to run after the dewar was filled, the TallBo dewar was first evacuated by a turbo pump to reduce contamination from residual gases.  It was then back-filled with gaseous argon.  The gaseous argon was next replaced with ultra-high-purity (UHP) LAr that passed through a molecular sieve and a copper filter.  This process took several days.  The volume of LAr in TallBo was approximately 460~liters. 

Once filling was complete, the TallBo dewar was sealed and subsequently maintained at a positive internal pressure of 8 psig to assure no contamination from the outside.  Gaseous argon from the ullage was recondensed to liquid argon with a liquid nitrogen condenser and then returned to the dewar. 


\subsection{Run Rates and Selection}
\label{sec:runRates}

The experiment consisted of recording the scintillation light from single cosmic muons traversing the LAr in the TallBo dewar defined by a four-fold coincidence trigger described in the Appendix.  

There were four periods of running, defined by the detector configuration read out.  The dates of these runs are given in Table~\ref{tab:experimentalRuns}.   The dewar was actually filled on 7/5/18 but problems with the new LAr filling system prevented data taking from beginning until 8/3/18.
\begin{table}[h]
  \begin{center}
    \caption{Experimental Runs }
    \vspace{0.2em}
    \label{tab:experimentalRuns}
    \begin{tabular}{| c |  c c c  c  |}
      \hline
      \hline
     & Tracks & Center & Run & Run  \\
      Run        &  Analyzed & Detector &  Start  &  End  \\  
      \hline
      
       1 & front & IU &  Aug 3, 2018 &  Aug 19, 2018 \\
       2 & front & IU  &  Aug 20, 2018 &  Aug 27, 2018  \\
      3& back & IU  & Aug 28, 2018 & Sept 9, 2018   \\
 \hline
drain/refill &&&&\\
\hline
       4& front & Fermi  &  Sept 18, 2018 & Oct18, 2018   \\
 
        \hline
      \hline
   \end{tabular}
  \end{center}
\end{table}
Run~1 and run~2 differed only by cable switches between monitor photosensors (Appendix).  
For run 3 the hodoscope scintillator paddles were switched so that back side tracks were read out.  This switch was made to ascertain whether front side and back side tracks gave similar results, as was expected.   

For run 4, the positions of the IU and Fermi PD modules were exchanged.  The purpose of this switch was to determine whether the geometric placement of the light guides affected the results.  There was a concern that the scintillation signal from cosmics would be affected by the position of the light guides with respect to the reflecting dewar walls.  To prepare for run 4, the dewar was first drained and the detectors warmed.  This process took two days.  Then over two days, the detctor frame was removed from the dewar, the light guide positions switched, and the frame remounted in the dewar.  The dewar was then pumped down and refilled.  The readout boards remained in their original positions to keep the light guide response independent of the board response.  In addition, the hodoscope scintillator paddles were switched back to read out front side tracks.  Draining and refilling the dewar was the point at which thermal cycling occurred.

The event statistics are given in Table~\ref{tab:runStats}.  Column [2] gives the total number of hours of run time for the runs.  The number of four-fold coincidence triggers per hour for each run are given in column [3].  Column [4] gives the number of four-fold coincidence triggers per hour in which there was one and only one hit on a PMT in each hodoscope module (``Single Track Rate'').  These single track events were the ones analyzed.  Table~\ref{tab:runStats} shows clearly that there was an abrupt change in the trigger rates between run 1 and run 2, run 3, and run 4.  This change in the rates is discussed below.
\begin{table}[h]
  \begin{center}
    \caption{Run Statistics}
    \vspace{0.2em}
    \label{tab:runStats}
    \begin{tabular}{| c |  c c c |}
      \hline
      \hline
        & Run & 4-fold  & Single Track \\
     Run         & Time  &  Coinc. Rate  &  Rate \\  
             &  [hr] &  [Hz]  &  [Hz]\\  
      \hline
      
       1 & 343 & 0.085  & 0.024 \\
       2 & 333 & 0.062 & 0.015  \\
       3 &  203 & 0.056 &0.013\\
\hline
drain/refill &&&\\
\hline
        4 & 612 & 0.064 &0.016 \\
 
        \hline
      \hline
   \end{tabular}
  \end{center}
\end{table}

\begin{figure}[h]
\centering
\includegraphics[width=2.7in,height = 2.4in]{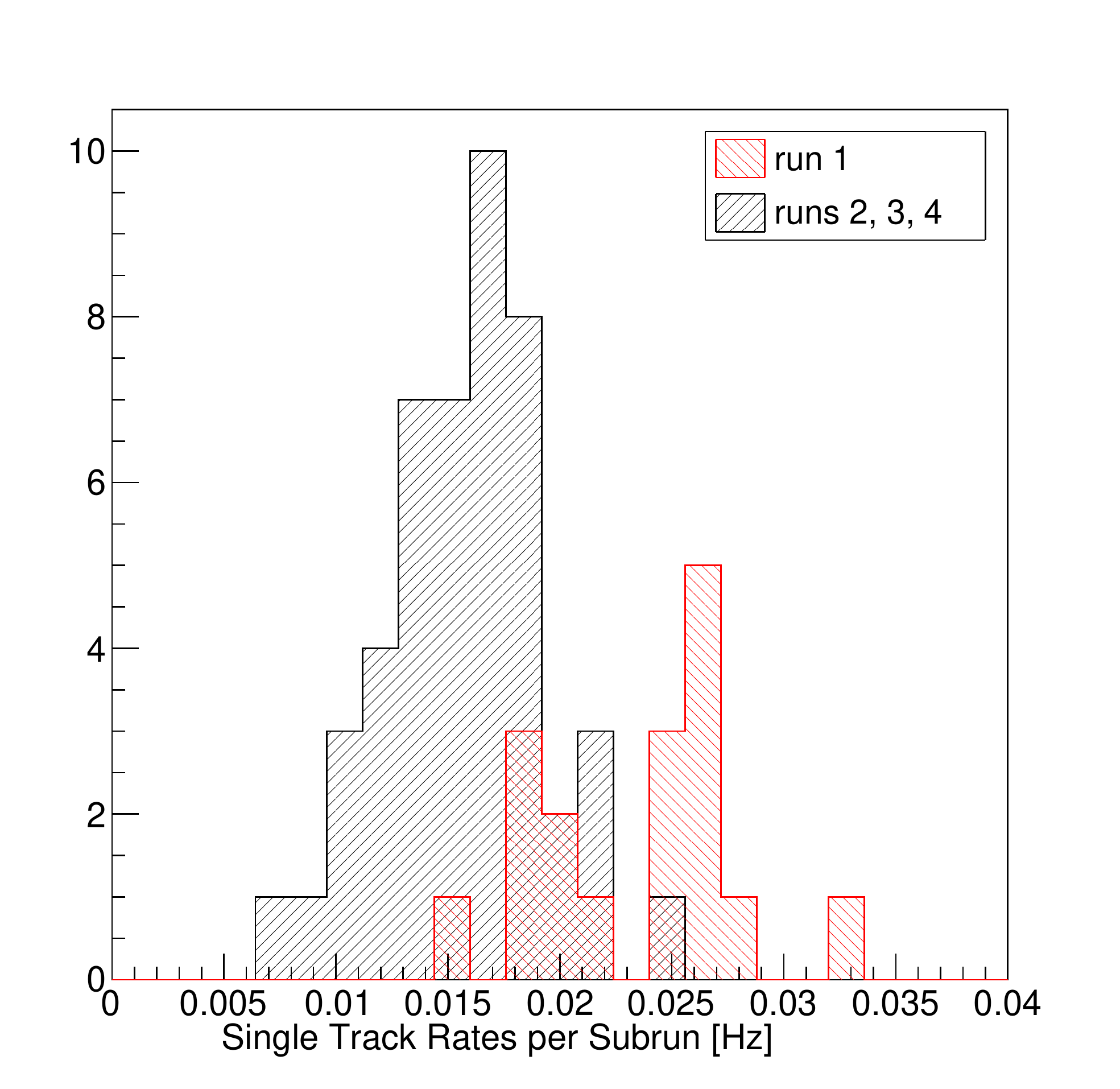}
\caption{The superposition of a histgram for the single track rates for the subruns in run 1 and a histogram for the single track rates for the subruns in runs 2, 3, and 4.}
\label{rateHistos}
\end{figure}

Each run in Table~\ref{tab:runStats} consisted of a series of 24-hour subruns.  Fig.~\ref{rateHistos} shows the superposition of a histogram for the single track rates for the subruns in run~1 and a histogram for the single track rates for the subruns in runs 2, 3, and 4.   
Fig.~\ref{rateHistos} shows that the individual subrun rates for run~1 were significantly higher than those for runs 2, 3, and 4.  The increase in the rates seen in Table~\ref{tab:runStats} is not the result of a few hot runs. There was nothing obviously different in the environment in PAB during run~1.  Since the rates change was seen after cable switches and a DAQ restart, it is likely something varied in the electronics.  But the actual cause is unknown.

\subsection{Contaminants}
\label{subsec:contaminants}

The contaminants that most affect LAr scintillation light are N$_2$, O$_2$, H$_2$O, and Xe, which can both decrease the argon transparency by absorbing scintillation light~\cite{bib:MITN2,bib:O2contamCross,bib:H2O,bib:H2OcontamCross,bib:neumeierXe} and quench scintillation light by collisional de-excitation of the Ar$^*_2$ dimer~\cite{bib:N2Contamination,bib:O2Contamination,bib:kubota}.  
The LAr delivered to TallBo  is UHP LAr from Airgas and is typically delivered with specified low levels of N$_2$, O$_2$, H$_2$O\footnote[2]{http://airgassgcatalog.com/catalog/P6-P7$\_$Argon$\_$Pure$\_$Gases.pdf}.  Airgas, however, does not specify the Xe contamination.  Since Xe is a valuable commercial gas, presumably Airgas filters almost all of it out to sell elsewhere.  But residual trace amounts are possibly present at unknown levels since no Xe measurement apparatus was available at TallBo during the experiment. 

The UHP LAr is pumped into TallBo through filters which are very effective at removing O$_2$ and H$_2$O.  Neither N$_2$ nor Xe is removed by the filters.  

The concentrations of N$_2$, O$_2$, and H$_2$O were monitored with commercial gas analyzers by Fermilab staff.  N$_2$ was monitored with an LDetek LD8000 trace impurity analyzer; O$_2$ was monitored with a Delta F Corporation DF-310$\epsilon$ process oxygen analyzer; and H$_2$O was monitored with a Tiger Optics M7000 HALO trace gas analyzer (manuals can be found in the Fermilab LArTPC DocDB).  The concentration of Xe could not be monitored. 

\begin{figure}[h!]
\centering
\includegraphics[width=2.4in,height = 2.1in]{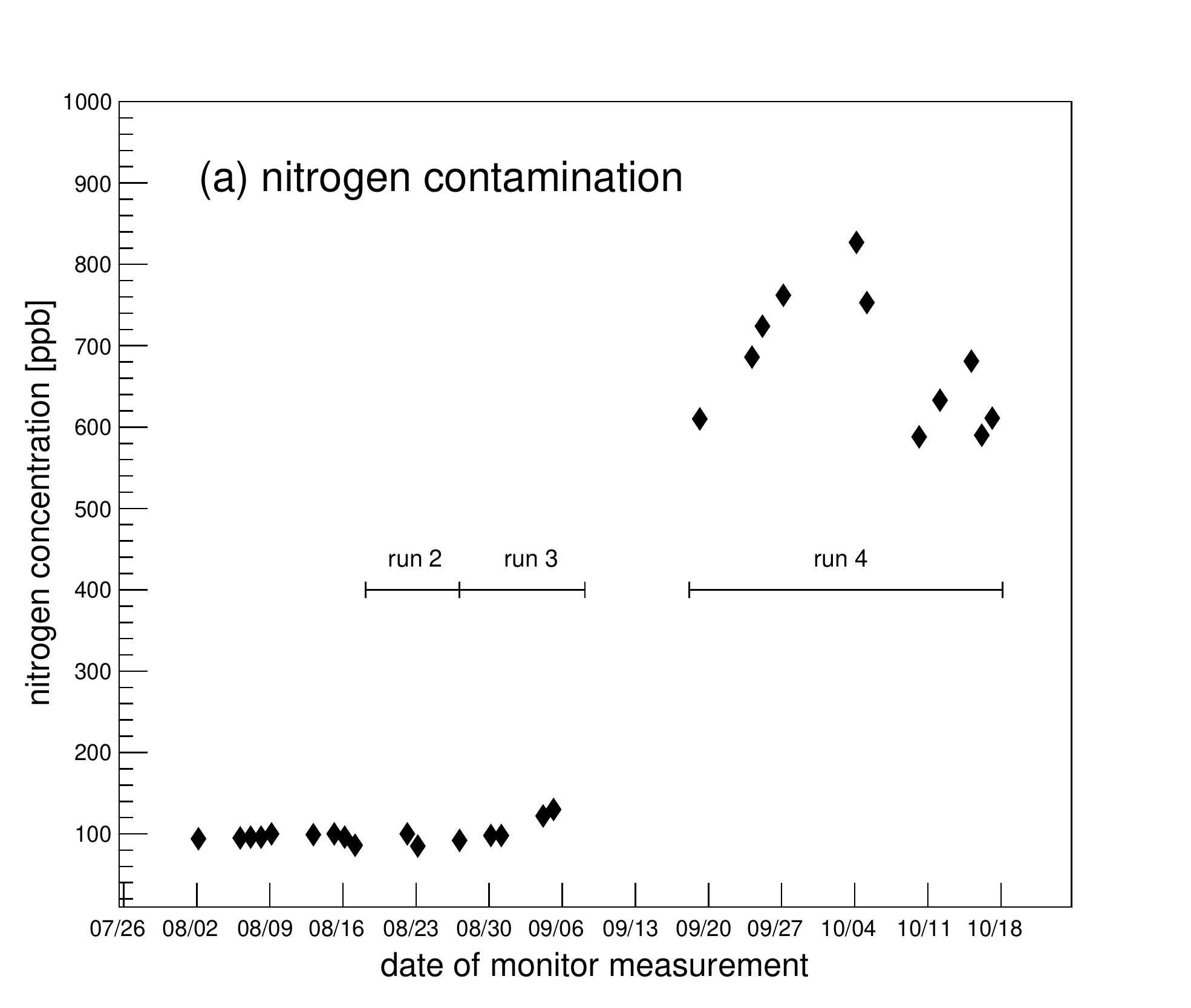}
\includegraphics[width=2.4in,height = 2.1in]{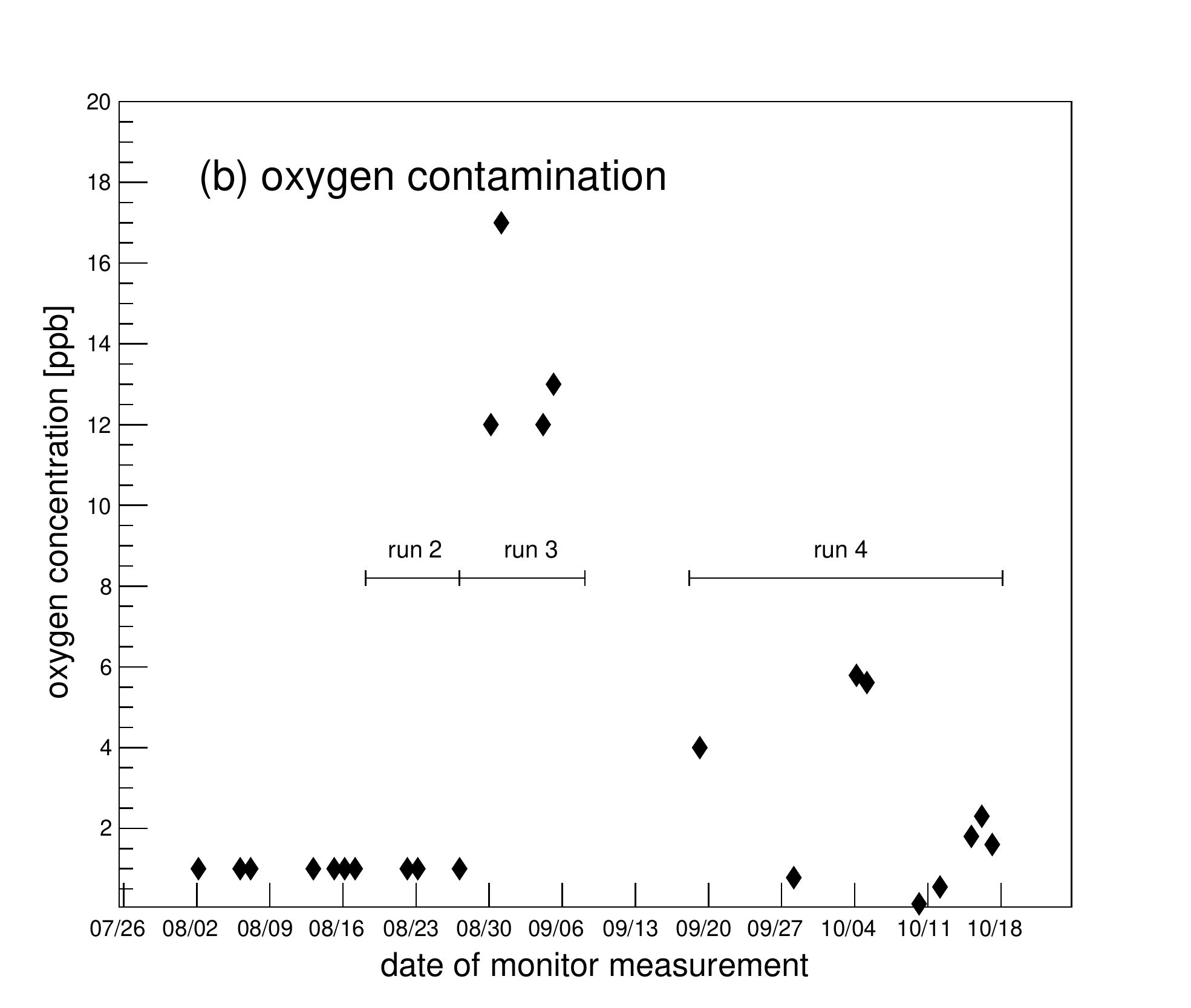}\\
\includegraphics[width=2.4in,height = 2.1in]{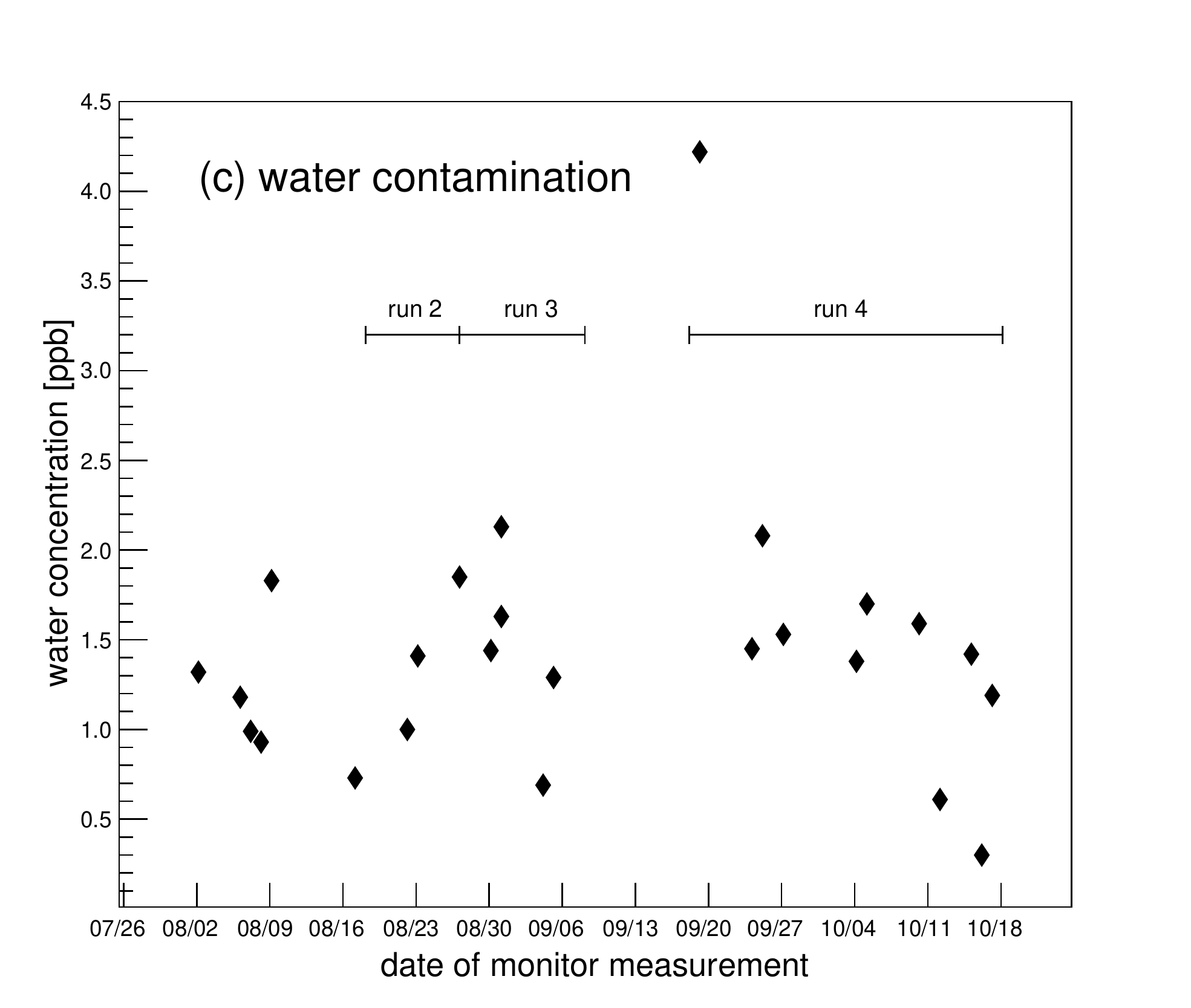}
\includegraphics[width=2.4in,height = 2.1in]{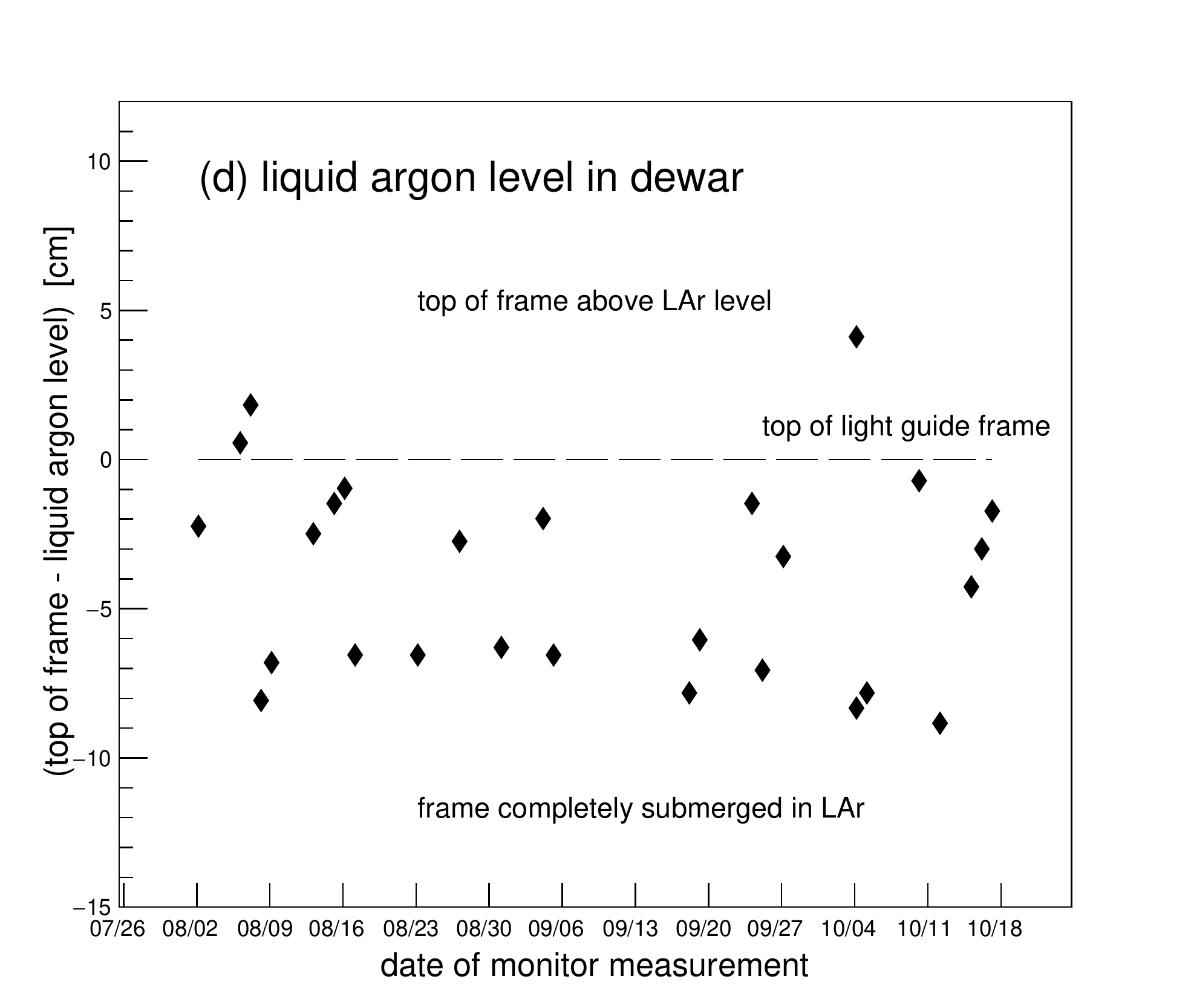}\\
\caption{The contaminants in the liquid argon and the LAr level in the dewar:  (a) The nitrogen contamination; (b) the oxygen contamination; (c) the H$_2$O contamination; and (d) the LAr level in the dewar.  Run dates are indicated on the figures.}
\label{contaminants}
\end{figure}

The concentrations of N$_2$, O$_2$, and H$_2$O during the experiment are shown in Fig.~\ref{contaminants}.  In addition, Fig.~\ref{contaminants} shows the liquid argon level relative to the top of the detector frame.  Only during brief excursions was the top of the frame above the liquid argon level.
The high N$_2$ readings during run 4 are likely due to the fact that the LAr was delivered with higher than typical N$_2$ contaminataion, although its concentration falls within Airgas UHP specifications.  The O$_2$ were high in run 3 and the filters were consequently regenerated after the dewar was drained in run 3.  The mean fractional number of contaminant molecules, $<\chi> =$, for runs 2, 3, and 4 computed from the measurements shown in Fig.~\ref{contaminants} are given in Table~\ref{tab:contaminants}.
\begin{table}[ht]
	\begin{center}
	\caption{Mean concentration $<\chi>$ of contaminants.}
	\vspace{0.2em}
	\label{tab:contaminants}
	\begin{tabular}{| cc c c|}
		\hline
		\hline
		 contaminant& run 2 & run 3 & run 4\\
		 & [ppb]& [ppb] & [ppb]\\
		\hline
                   N$_2$ & 92& 109&679\\
                   O$_2$ & 1.4& 14.2&1.7\\
                   H$_2$O & 1.4 & 1.4 & 1.3\\
 		\hline
		\hline
	\end{tabular}
	\end{center}
 \end{table}
The one outlier H$_2$O measurement in run~4 is well outside the range of all others and is thought to be due to a glitch in the monitor at an LAr topoff.  This level of H$_2$O was not repeated during the run and was not used in computing $<\chi>$ for H$_2$O.

\FloatBarrier
\section{TallBo Simulation}
\label{sect:simulation}

The track simulation follows closely that described in~\cite{bib:howard}.  Briefly, in this simulation cosmic muons are treated as traveling along straight paths with end points fixed at the centers of the two triggered PMTs in the hodoscopes on either side of the dewar.  It is assumed that the cosmic muons are minimum ionizing particles that generate 40,000 photons/MeV~\cite{bib:howard,bib:scintYield2} or $8.42 \times 10^4$ photons/cm along their tracks.  The starting positions of the scintillation photons were distributed uniformly along the track segment passing through the LAr volume and the photons' momentum vectors were distributed uniformly in solid angle.  Photons were tracked along straight line paths until they intersected with either an IU or Fermi photon detector or were lost.  Along their tracks, photons could undergo a Rayleigh scattering or reflect off the walls of the dewar.  Every track in the TallBo data set that struck two single hodoscope PMTs and passed data selection cuts was simulated. 

%
%
\section{Results}
\label{sect:results}

\subsection{Data Analysis}
\label{dataAnalysis}

All events selected for analysis were four-fold coincidence triggers with at least one hit in each hodoscope module on opposite sides of the dewar.  Each trigger was further required to have one and only one hit on a PMT in each hodoscope module and hits in its adjacent scintillator plane.  If a straight line track between the PMT centers crossed the detector plane, the event was discarded.  If a track was on the back side for runs 2 and 4 or the track was on the front side for run 3, the event was discarded.  The sample of events selected was assumed to be dominated by single minimizing muons passing through the liquid argon.  Analysis cuts are further described in $\S$\ref{AnalysisAppendix}

The signal from the selected waveforms was integrated.  The start point for the integration is the trigger at 2 $\mu$s.  The end point for the integration, however, depends on whether the waveform was read out by a passive board or an active board.  As shown in Fig.~\ref{boardWFsAppendix}, the waveforms from the passively ganged boards and the actively ganged boards are distinctly different.  For the passively ganged boards, the integral was extended out to 10.7 $\mu$s to collect all the charge information.   It is not so clear, however, how far to extend the integral for the active boards.  The main concern was whether to include in the integral the piece of the waveform after the waveform overshoots the baseline.  
In this analysis the choice was made to integrate the waveforms from AGB1 and AGB2 only out to the overshoot.  This choice is explained in $\S$\ref{AnalysisAppendix}..

The integrated waveforms from the passive and active boards on the IU light guide and the Fermi light guide were put into separate histograms for runs 2, 3, and 4.  Examples of these histograms for the passive boards on the IU light guide in runs 2 and 4 are shown in Fig.~\ref{IUrun2-4Appendix}.
%
Two cuts were made on the histograms.  (1) Integrated waveforms with very low ADC counts were considered noise and cut.  (2) As seen in Fig.~\ref{IUrun2-4Appendix}, there is a large peak and a long, high energy tail in the histograms of the single track events.  The distribution of events around the peak (``bump'') was associated with single, minimum ionizing cosmic muons.  The long tails are likely a mix of high energy muons and noise events.  These events were also cut from the analysis.  For each run in the analysis, the cuts were made independently for the passive boards and the active boards.  For each run the waveform samples for PGB1 and PGB2 included the same tracks, and the waveform samples for AGB1 and AGB2 included the same tracks.  The passive and active samples, however, were not the same.

The results of the analysis are given in Table~\ref{tab:Results}.  
\begin{table}[ht]
\begin{center}
	\caption{Results.}
	\vspace{0.2em}
	\label{tab:Results}
	\begin{tabular}{| c | c c c c c |}
		\hline
		\hline
&readout & $\#$tracks &	 bump integral &  geometry  & [ADC]/track\\
& board & & [ADC] & correction & \\  
	\hline
\underline{IU light guide} & \underline{passive}  & &  & & \\  
run2 & PGB1 & 2,391 & $1.18\times10^9$ &1.00 & $4.94\times10^5$\\  
run3 & PGB1 & 2,322 & $1.14\times10^9$ &1.00 & $4.91\times10^5$\\
run4 & PGB2& 10,160 & $3.03\times10^9$ &1.04 & $2.98\times10^5$\\
&\underline{active} &&&&\\
run2 & AGB1 & 683 & $2.63\times10^8$ &1.00 & $3.54\times10^5$\\  
run3 & AGB1 & 997 & $1.69\times10^8$ &1.00 & $3.70\times10^5$\\
run4 & AGB2 & 1,217 & $2.86\times10^8$ &1.04 & $2.35\times10^5$\\
\hline
\underline{Fermi light guide} & \underline{passive}  & &  & & \\  
run2 & PGB2 & 2,380 & $6.06\times10^8$ &1.04 & $2.65\times10^5$\\  
run3 & PGB2 & 2,249 & $6.35\times10^8$ &1.02 & $2.82\times10^5$\\
run4 & PGB1& 10,187 & $2.10\times10^9$ &1.00 & $1.98\times10^5$\\
&\underline{active} &&&&\\
run2 & AGB2 & 728 & $1.67\times10^8$ &1.04 & $2.29\times10^5$\\  
run3 & AGB2 & 886 & $9.79\times10^7$ &1.03 & $2.30\times10^5$\\
run4 & AGB1 & 2,311 & $5.71\times10^8$ &1.00 & $2.47\times10^5$\\
	\hline
	\end{tabular}
	\end{center}
\end{table}
The number of tracks in the muon bump for each readout board in each run is given in column [3].  Although the track sample for the passive boards in each run and the track sample for the active boards in each run are the same, the number of tracks in each sample used in the analysis differs somewhat as a result of the cut on the high energy tail affecting the two modules somewhat differently.  Further, it is apparent that there are significantly fewer tracks in the active board samples than in the passive board samples.  This is the result of cutting events with standard deviations $>$2, which impacts the waveforms in the active board sample more strongly, as shown in Fig.~\ref{SDrun2Appendix}.  Column [4] gives the total number of ADC counts in the waveforms in the bump.  

There was a concern that the differing proximity of the light guides to the reflecting dewar walls could systematically affect the number of scintillation photons striking the light guides.  The TallBo simulation was used to address this issue.  Every track in the TallBo data set that passed analysis cuts was simulated 10 times.  For each of these 10 simulated tracks, the positions of all photons that stike either the IU or Fermi PD modules were collected into a histogram.  The resulting 10 histograms for each PD were then averaged into a single histogram to determine the expected number of photons striking either the IU or Fermi PD the light guide for that track.  The individual track histograms were then summed for all tracks in the passive or active waveform samples in each run to find the number of scintillation photons striking either the IU or Fermi PD.  The ratio of the number of photons striking the PD modules, or the correction for the position of the light guides in the dewar for each run (normalized to PBG1 or AGB1) is given in column [5] of Table~\ref{tab:Results}.  Finally column [6] gives the mean $\#$ of ADC counts per track in the sample corrected for the position of the light guides in the dewar.

There are two results inferred from column (6) of Table~\ref{tab:Results}.  First, the switch from the front side tracks in run~2 to the back side tracks in run~3 had little effect on the mean $\#$ of ADC counts per track.  This was the expected result because there was no change in the readout boards and the light guides see scintillation photons symmetrically from both sides.  The mean variation in the $\#$ of ADC counts per track between the front side and the back side is $\approx$4$\%$.  Since the track trajectories in the samples of front side and back side tracks differ somewhat, the match is not exact.  The efficiency of the TPB plates (IU) or coating (Fermi) are also expected to be somewhat different.  

On the other hand, there was no clear expectation on how the mean $\#$ of ADC counts per track would differ from runs~2 and 3 to run 4, after the dewar was drained and refilled.  Table~\ref{tab:Results}, however, shows that there was a significant drop-off in the $\#$ of ADC counts per track after the light guides were exchanged 
for the IU and Fermi technologies when read out passively and for the IU technology when read out actively.  
Calculating the percent change as 

$(<\rm{run~2, run~3}> -~\rm{run~4})/[(<\rm{run~2, run~3}> +~\rm{run~4})/2] \times 100$ 

\noindent shows that the 
the percent change in the $\#$ of ADC counts per track fell by $49\%$ for the IU light guide read out passively, by $32\%$ for the Fermi light guide read out passively, and by $43\%$ for the IU light guide read out actively.  Only the Fermi light guide read out actively does not follow this trend, showing no significant fall-off.  

To test whether the results of Table~\ref{tab:Results} are robust, the analysis was repeated with each data set broken into two.  The same analysis routines were used with the same cuts.  For passive boards in runs~2 and 3, the ($\#$ ADC counts)/track for both pieces remain close to their values for the whole run.  For the active boards, where there are fewer tracks, the results show more variation but they remain consistent with Table~\ref{tab:Results}.  
For both pieces of the run~4 data set for the active board on the Fermi light guide, no drop-off is seen, consistent also with Table~\ref{tab:Results}.  

There is one additional result that can be deduced from column (6) in Table~\ref{tab:Results}.  Comparing the $\#$ of ADC counts per track for the IU and Fermi light guides for runs~2 and 3, before the dewar was drained, there is evidence that the IU light guide technology has a higher efficiency than the Fermi technology.  An estimate of the relative efficencies of the IU and Fermi technologies is best obtained from the passive boards, calculated using the ratio of the mean $\#$ of ADC counts in run 2 and run 3 for the two technologies.  This estimate implies that the IU technology is $\sim$16$\%$ more efficient than the Fermi technology.

\section{Discussion}
\label{sect:discussion}

There are several possible explanations for the significant drop-off in the efficiency of the IU and Fermi light guides after refilling the dewar for run~4. 

%

\subsection{Contamination}

One likely explanation for the drop-off seen in run 4 was the absorption or quenching of the scintillation light by contaminants.  


\subsubsection{Absorption}

The probability that a scintillation photon is absorbed by a contaminant along its path is given by $P_{abs} = A \exp{(-L/\lambda_{abs})}$, where $A$ is a normalization constant, $L$ is the track length to the intersection, and $\lambda_{abs}$ is the absoption length~\cite{bib:howard}.   The absorption length is given by $\lambda_{abs} = 1/n_c\sigma = 1/[<\chi> \cdot n_{LAr} \cdot \sigma],$
where $n_c =$ the number density of contaminants, $\sigma$ is the cross section for absorption of 128~nm VUV photons, $<\chi> =$ mean fractional number of contaminant molecules, and $ n_{LAr}$ = the number density of LAr atoms/cm$^3$  = (1.396~g/cm$^3$)/(39.948~g/mol) $\times \, (6.02 \times 10^{23}$~atoms/mol)~\cite{bib:PDG} $ = 2.1037\times 10^{22}$ LAr atoms/cm$^3$.  The absorption lengths of N$_2$, O$_2$, and H$_2$O for the three runs are given in Table~\ref{tab:abslength}.  
\begin{table}[ht]
	\begin{center}
	\caption{Absorption lengths of N$_2$, O$_2$, and H$_2$O in the 3 runs.}
	\vspace{0.2em}
	\label{tab:abslength}
	\begin{tabular}{| c  c  c  c c c|}
		\hline
		\hline
		contaminant& $\sigma$ & ref. &\multicolumn{3}{c|}{$\lambda_{abs}$~[m]}\\
		 & [cm$^2$] & &run 2 & run 3 & run 4\\
		\hline
                  N$_2$ & $7.1 \times 10^{-21}$ & \cite{bib:MITN2}& $7.9 \times 10^2$ & $6.1 \times 10^2$ & $9.9 \times 10^1$\\
                  O$_2$ & $2.8 \times 10^{-19}$ & \cite{bib:O2contamCross}& $1.2 \times 10^3$ & $1.2 \times 10^2$ & $1.0 \times 10^3$\\
                  H$_2$O& $8.0 \times 10^{-18}$ & \cite{bib:H2OcontamCross}& $4.2 \times 10^1$& $4.2 \times 10^1$& $4.6\times 10^1$\\
		\hline
		\hline
	\end{tabular}
	\end{center}
 \end{table}

The absorption lengths in Table~\ref{tab:abslength} and the the distribution of photon path lengths from the TallBo simulation of the sample tracks in $\S$\ref{sect:simulation} were used to compute the importance of absorption by N$_2$, O$_2$, and H$_2$O.  First a photon path length was drawn from the distribution of all photon path lengths simulated for that run.  Three absorption lengths for that photon for the three contaminants were drawn as exponential deviates using ROOT's TRandom3 package and $\lambda_{abs}$ from in Table~\ref{tab:abslength}.  If any of the 3 simulated absorption lengths were shorter than the photon path length, the photon was assumed lost to absorption before it reached the detectors.  The calculation was repeated $10^7$ times for each run.  The probability of absorption ${\bf P}_{abs}$ by any of the three contaminants in each run are given in Table~\ref{tab:absorpton}.
\begin{table}[ht]
	\begin{center}
	\caption{Photon absorption probability by contaminants in Table~\ref{tab:contaminants}.}
	\vspace{0.2em}
	\label{tab:absorpton}
	\begin{tabular}{| c c c c|}
		\hline
		\hline
		 absorption probability& run 2& run 3 & run 4\\
 		\hline
		~${\bf P}_{abs}$ & 0.009 & 0.009 & 0.011 \\
		\hline
		\hline
	\end{tabular}
	\end{center}
 \end{table}
These calculations show that the level of N$_2$, O$_2$, and H$_2$O contaminatopn in TallBo shown in Table~\ref{tab:contaminants} have only a minor effect on absorption the scintillation light before it reaches the detectors.  Also, since the absorption probability is approximately equal during all 3 runs, absorption by these contaminants did not introduce run-by-run differences in the results.

These results are consistent with previous studies.  Absorption by N$_2$ is not expected to affect the scintillation light at the concentrations in Table~\ref{tab:contaminants}~\cite{bib:MITN2,bib:N2Contamination}. 
At concentrations $<$100~ppb, absorption by O$_2$ contamination on scintillation light in LAr are expected to be negligible~\cite{bib:O2Contamination}.  Absorption by H$_2$O is not well studied.  
However, studies in gaseous argon suggest the H$_2$O concentration at this level does not affect our results~\cite{bib:H2O}.  Simulations confirm these expectations.

Absorption by unknown trace amounts of Xe, however, is more problematical.  The results of Neumeier {\it et al.}\cite{bib:neumeierXe} show that even trace contamination of Xe down to 100 ppb can absorb a significant fraction of the 128~nm scintillation light from LAr.  This absorbed energy is then reradiated by the Xe at 174~nm on a time scale shorter than pure LAr~\cite{bib:Wahl}.  However, no more energy can be radiated at 174~nm than was absorbed at 128~nm.  Since the detectors in this experiment are sensitive to 174~nm photons, the relative scintillation signal strength in runs~2 and 3 compared with run~4 due to differences in the trace contamination of Xe depends on many factors, including the unknown contamination concentration of Xe, the window used to integrate the waveforms, the relative efficiency of TPB absorption at 128~nm and 174~nm, and the relative efficiency of the MPPCs at 128~nm and 174~nm.  Consequently, it remains a possibility that unknown differences in the Xe contamination in runs~2 and 3 compared with run 4 could account for the drop-off observed.  

\subsubsection{Quenching}

Using the formalism of \cite{bib:N2Contamination} and \cite{bib:O2Contamination}, the effects of quenching by N$_2$ and O$_2$ can be computed.  There is no information on quenching by H$_2$O, but the fact that its level of concentration does not change during the experiment (Table~\ref{tab:contaminants}) means that quenching by H$_2$O cannot explain the drop-off.

The fraction of Ar$^*_2$ dimers that survive quenching is characterized by the quenching fraction, $Q_F = A'_S + A'_I + A'_T$, where $A'_j$ are the quenched amplitudes for the short, intermediate, and triplet decay modes of the Ar$^*_2$ dimer.  Here $A'_j = A_j/(1 + \tau_j  k_Q [N_2, O_2])$, where the $A_j$ are the normalized unquenched amplitudes.  For  N$_2$ , $A_S = 18.8\%$ for the singlet state,  $A_I = 7.4\%$ for the intermediate state, and $A_T = 73.8\%$ for the triplet state; for O$_2$, $A_S = 24.6\%$ and $A_T = 74.8\%$.  The $\tau_j$ are the lifetimes of the states.   For  N$_2$, $\tau_S =$ 4.9~ns, $\tau_I =$ 34.0~ns, $\tau_T =$ 1260~ns; for O$_2$, $\tau_S =$ 5~ns and $\tau_T=$ 1210~ns.  The quenching rate constants are $k_Q(N_2) = 0.11$~ppm$^{-1}$~$\mu$s$^{-1}$ and 
$k_Q(O_2) = 0.54$~ppm$^{-1}$~$\mu$s$^{-1}$~\cite{bib:N2Contamination,bib:O2Contamination}.  $[N_2, O_2])$ 
are the concentrations of the nitrogen or oxygen contaminants.   With the contaminant concentrations given in Table~\ref{tab:contaminants}, the percentage of scintillation light lost to quenching, $100 - Q_F$, in each of the runs is given in Table~\ref{tab:quenchingFraction}.
\begin{table}[ht]
	\begin{center}
	\caption{Percentage of scintillation light lost to quenching.}
	\vspace{0.2em}
	\label{tab:quenchingFraction}
	\begin{tabular}{| c  c c c c|}
		\hline
		\hline
		contaminant& ref. & \multicolumn{3}{c|}{(100. - Q$_{F})~[\%]$}\\
		 && run 2& run 3 & run 4\\
		\hline
                  N$_2$ & \cite{bib:N2Contamination} & 0.9& 1.1& 6.4\\
                  O$_2$ & \cite{bib:O2Contamination} & 0.07 & 0.7&0.08\\
		\hline
		\hline
	\end{tabular}
	\end{center}
 \end{table}
What matters is the difference between the quenching fraction in runs 2 and 3 and the quenching fraction in run 4.  Table~\ref{tab:quenchingFraction} gives $\Delta$(100 - Q$_F$) = $\sim$5.4\% for N$_2$ between the two fills and is negligible for O$_2$.   Quenching by N$_2$, O$_2$, or H$_2$O cannot explain the degradation in the light yield in Table~\ref{tab:Results}. 

Quenching by Xe  also cannot explain the-drop off in signal in run~4.  Assuming conservatively that all scintillation light is from the triplet state with $\tau_T =$ 22~ns~\cite{bib:Wahl} and an increase in the Xe contamination of 1 ppm in run~4, larger than  might reasonably be expected, would only reduce the scintillation light by 1.5$\%$.


\subsubsection{Other components of scintillation light}

There is evidence of additional components of scintillation light from O$_2$ contamination at 200~nm and 577~nm~\cite{bib:LArScint,bib:Belov,bib:Johnson}.  From Table~\ref{tab:contaminants}, the mean O$_2$ contamination is approximately equal in runs~2 and 4, but there is a significant drop-off in scintillation light in run~4, implying that the extra scintillation light from O$_2$ does not play a role.  Further, there is a significant rise in the O$_2$ contamination in run~3 from run~2 without any rise in the light signal from run~2, again implying that O$_2$ contamination was not a factor.

There can be additional scintillation light from H$_2$O contamination if the molecule is split and excited OH forms~\cite{bib:morozov}.  However, the H$_2$O contamination does not vary during the experiment (Table~\ref{tab:contaminants}), so this is unlikely to be a factor.

\subsection{MPPC Response}
\label{MPPCresponse}

\begin{figure}[t!]
\centering
\includegraphics[width=2.3in,height = 1.9in]{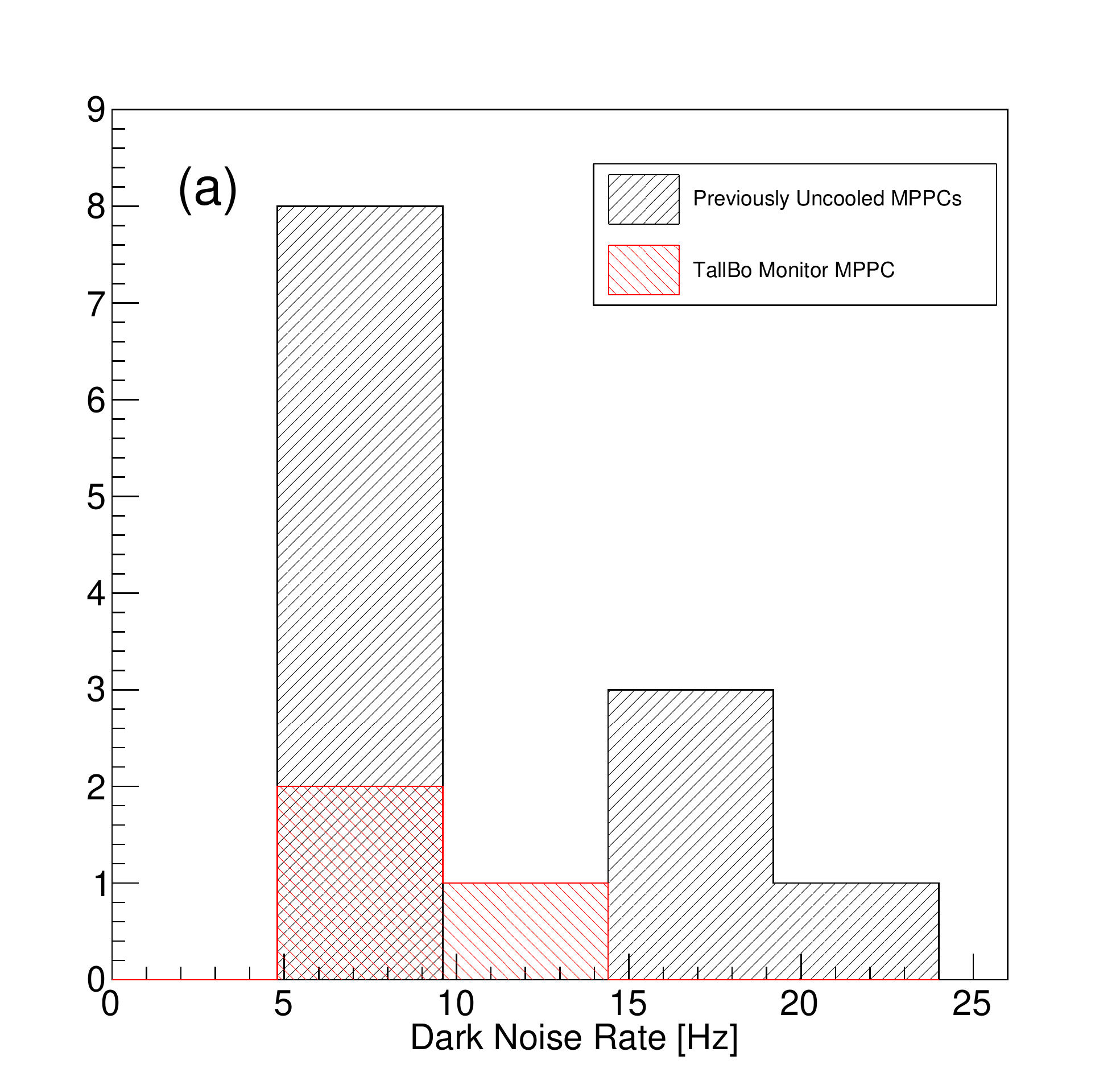}
\includegraphics[width=2.3in,height = 1.9in]{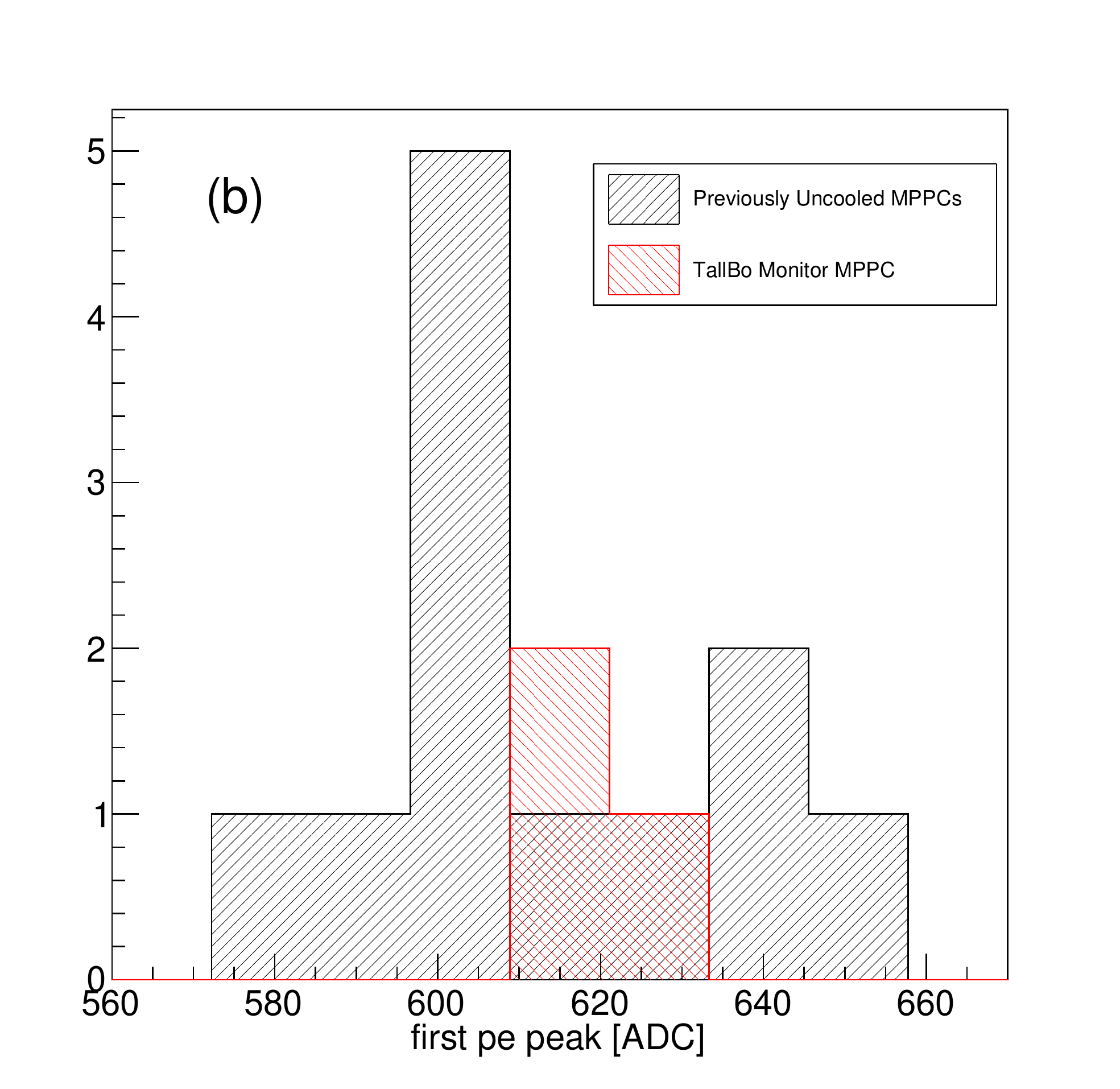}\\
\includegraphics[width=2.3in,height = 1.9in]{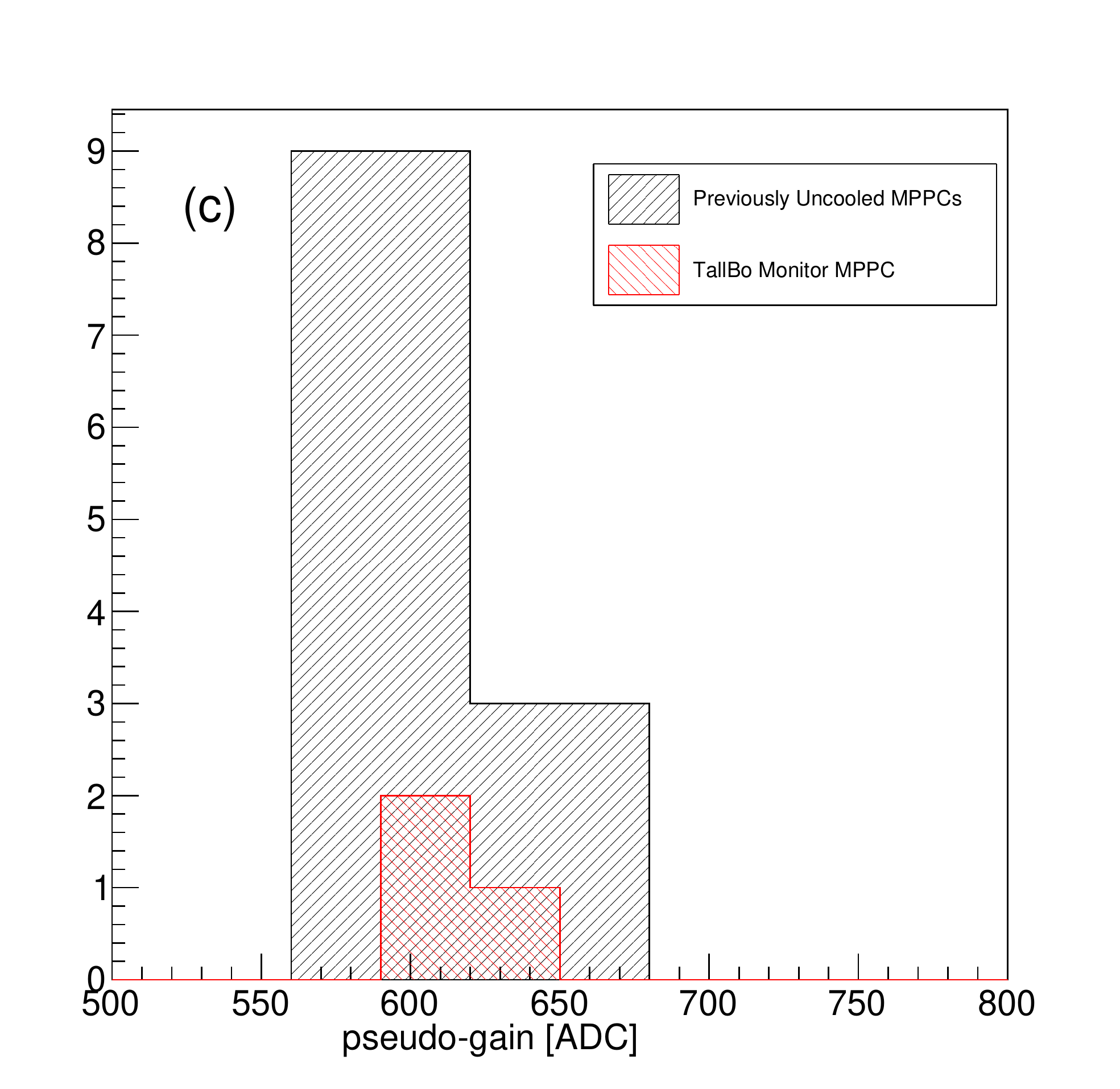}
\includegraphics[width=2.3in,height = 1.9in]{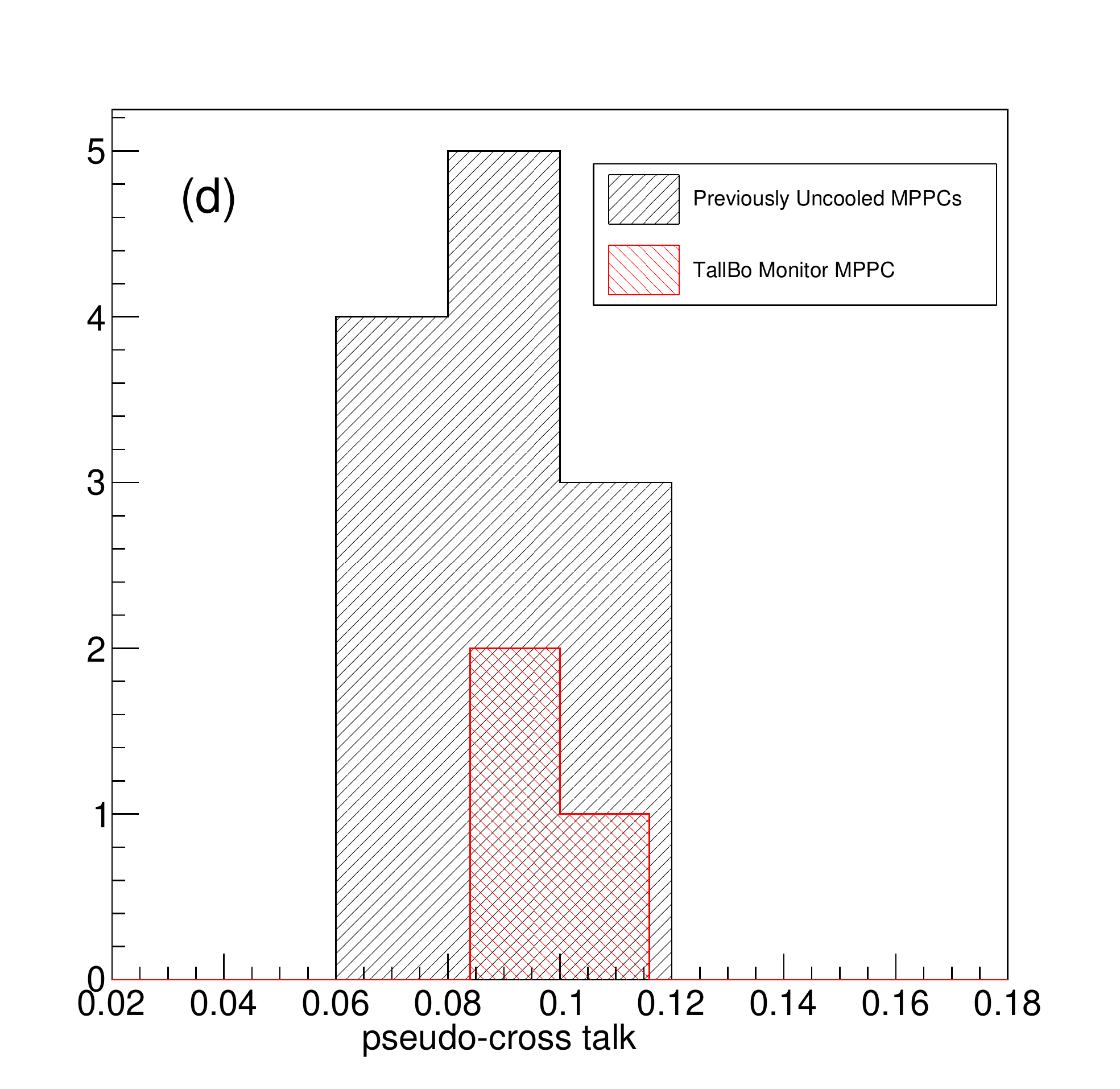}\\
\caption{Laboratory measurements of the the performance characteristics of 12 previously uncooled Hamamatsu MPPCs immersed in LN2 biased at 44.5~V.  Superposed on these histograms are 3 independent laboratory measurements of the TallBo monitor MPPC (c.f., Fig.~\ref{exptLayout}).  (a) The dark noise is the sum of the ADC counts divided by the data acquisition time; (b) the number of ADC counts at the peak of the first p.e. peak; (c) the pseudo-gain or the number of ADC counts between the first and second p.e. peaks; and (d) the pseudo-cross talk determined as the ratio of the number of ADC counts in the peak of the second p.e. peak to the number of ADC counts in the peak of the first p.e. peak.}
\label{MPPCcharacteristicsComp}
\end{figure}

A second possibility is that the MPPC response degraded due to the refilling of the dewar prior to run~4.  Fig.~\ref{MPPCcharacteristicsComp} shows laboratory measurements of the the performance characteristics, -- dark noise, position of the first p.e. peak, the pseudo-gain, and the pseudo-cross talk -- of 12 previously uncooled Hamamatsu MPPCs immersed in LN2.   The MPPCs were biased at 44.5~V. 
The description of these characteristics are described with Fig.~\ref{MPPCcharacteristicsAppendix}.
Superposed are 3 independent laboratory measurements of these characteristics of the TallBo monitor MPPC (c.f., Fig.~\ref{exptLayout}) made after the experiment was completed.  The characteristics of the monitor MPPC are consistent with the characteristics of the test devices that had never been cooled implying that the monitor MPPC was not significantly affected by the environmental conditions it experienced during the experiment.

Fig.~\ref{compMPPCsAppendix} in the Appendix is further evidence that this explanation is also unlikely.  Fig.~\ref{compMPPCsAppendix} shows the comparison of the dark spectrum of a previously uncooled MPPC in LN2 biased at 44.5~V (cf., Fig.~\ref{testMPPCAppendix}) compared with the dark spectrum of the monitor MPPC also biased at 44.5~V.  
Both MPPCs demonstrate very similar dark spectra.  
In addition, the 48 individual MPPCs in the 4 readout boards were removed once the experirment was completed and tested electrically by measuring their resistance and continuity.  Their properties were similar to what they were before the experiment.  



\subsection{Readout Boards}
\label{readoutBoards}
A third possibility is that the readout boards degraded after the refill.  
Fig~\ref{passiveBoardAppendix} suggests this is unlikely for the passive readout boards.  This figure shows the baseline subtracted mean of 25 waveforms from single track muon events crossing TallBo as read out by a passive ganging board in runs 2 (front), run 3 (back), and run 4 (front-reversed).   
Since the positions of the light guides were exchanged from run 3 to run 4, but the readout boards remained in place, the waveforms are labeled with the readout boards that collected them.  The waveforms were multiplied by the geometry correction given in Table~\ref{tab:Results}.  The 25 waveforms used in computing the mean were selected near the peak of the muon bump (Fig.~\ref{IUrun2-4Appendix}).
The reduced signal strength responsible for the drop-off in efficiency is seen in the depth of the waveform response.
This is unlikey to be result of a degradation in the individual MPPC response. ($\S$\ref{MPPCresponse}).
Fig.~\ref{passiveBoardAppendix} shows the the waveforms from the boards are qualitatively similar, both before and after the refill, and that the exponential tails remain comparable, suggesting that the board response has also not degraded.  Most likely it is the light guides responsible for the degradation seen in the passively ganged technologies.

The light guides read out by the active boards show more complicated behavior.  
Fig~\ref{activeBoardAppendix} shows the baseline subtracted mean of 25 waveforms from single track muon events crossing TallBo as read out by an active ganging board in runs 2 (front), run 3 (back), and run 4 (front-reversed).  
For the IU light guide there is again reduced signal strength as seen in the depth of the waveform response.
Since the waveforms all have similar structure before and after the refill, with the overshoot occurring at approximately the same point on the waveform, this suggests that the board response has not degraded.  

For the Fermi light guide the situation is different.  Although the waveforms have very similar forms in the runs, with the overshoot occurring at approximately the same point on the waveform, signal strength as seen in the depth of the waveform response has not fallen off.  The most likely explanation for this behavior is that the response of AGB1 is greater than the response of AGB2.  
Lab tests of this hypothesis on the AGB boards at IU once the experiment was completed were not successful.  
Reflections off the stainless steel sides of the small test dewar made the results dependent on the exact placement of the boards in the dewar and it proved impossible to reproduce the exact position of the boards in the dewar from test to test.  

\subsection{Difference in Track Length}

\begin{figure}[h!]
\centering
\includegraphics[width=2.4in,height = 1.8in]{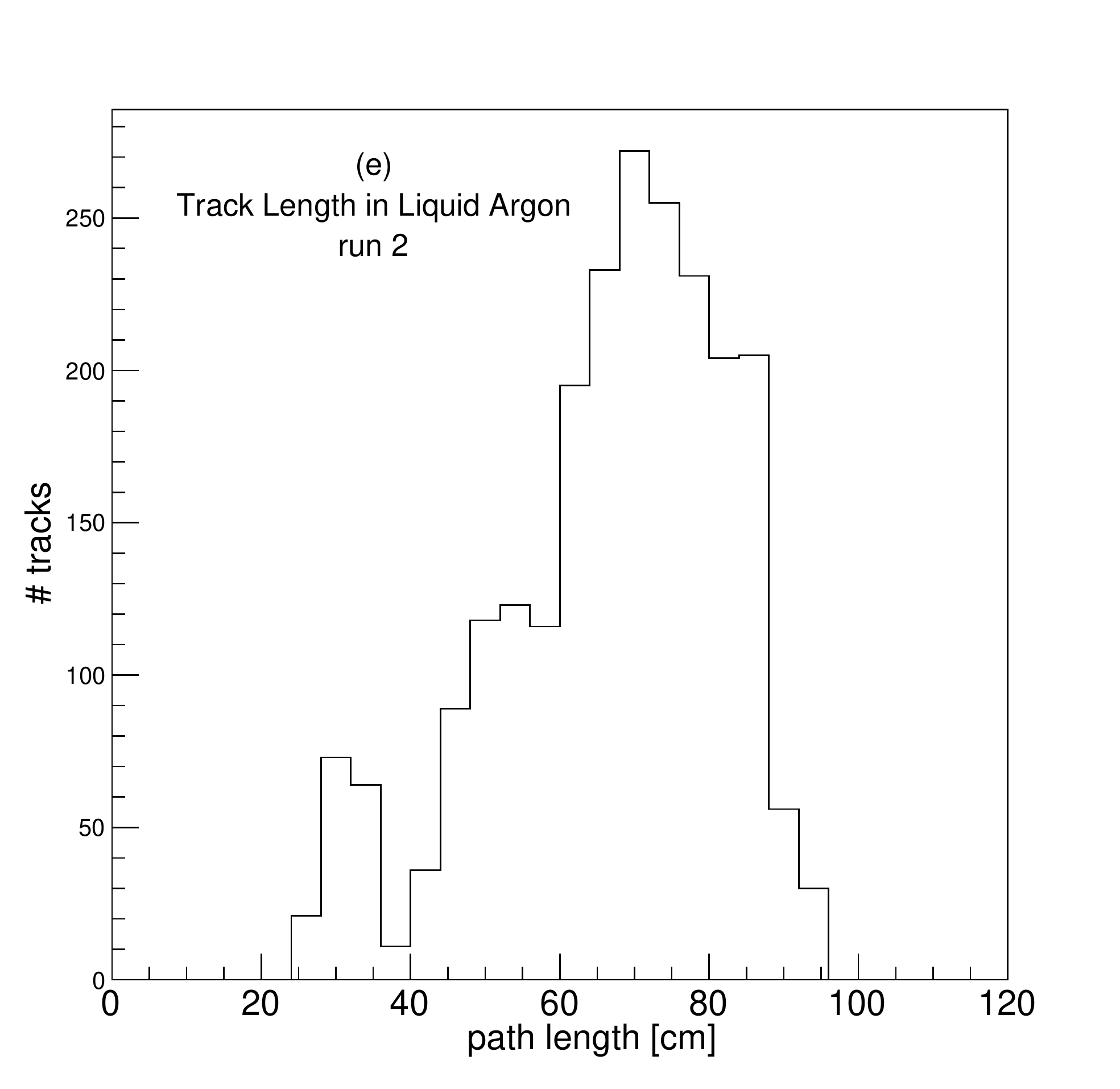}
\includegraphics[width=2.4in,height = 1.8in]{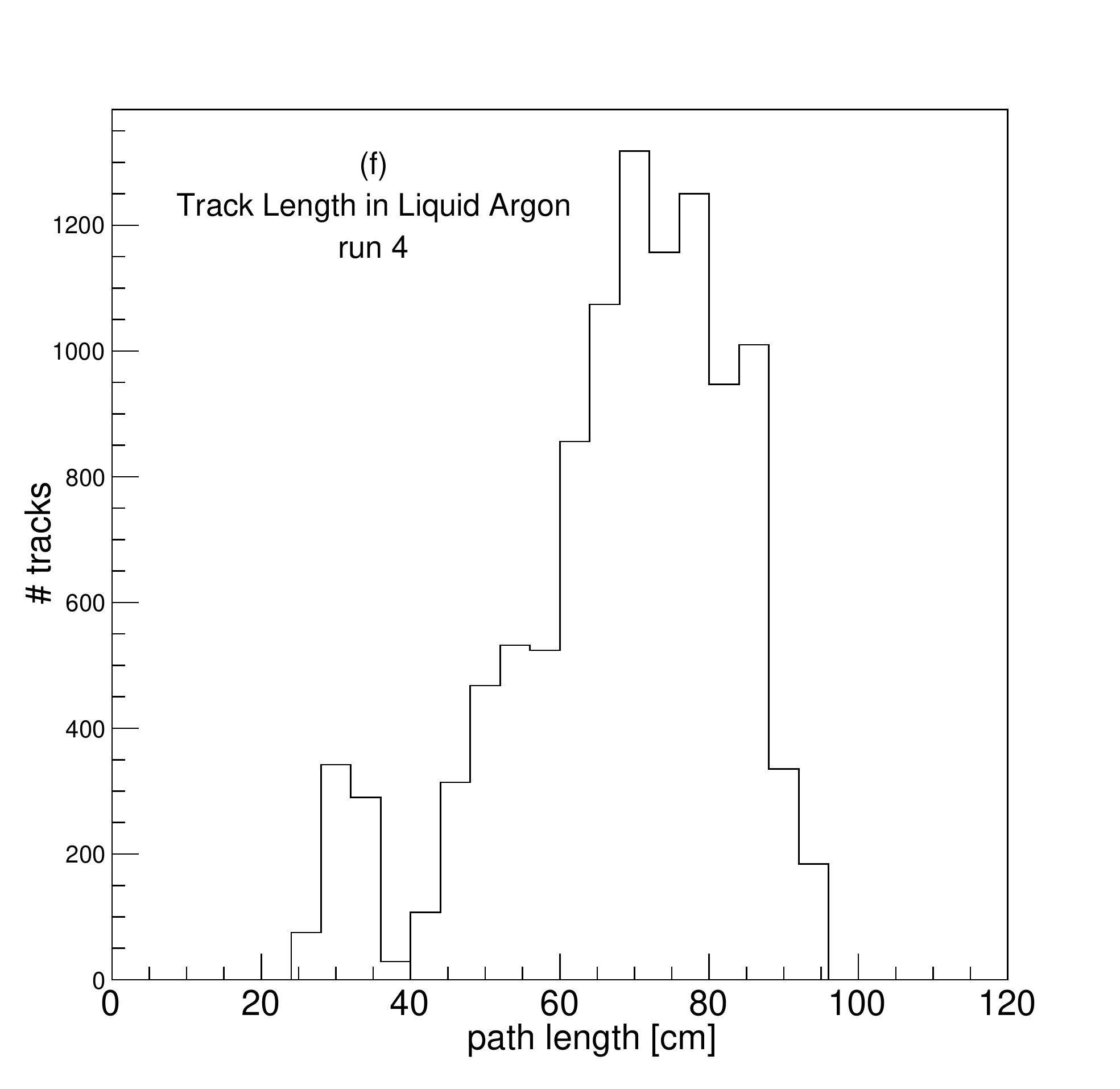}\\
\caption{(a) the track length distribution in run 2; and (b) the track length distribution in run 4.}
\label{Distributions}
\end{figure}

The drop-off in signal from run 2 to run 4 is also unlikely to be due to a difference in the track length distributions in LAr of the tracks in the two runs.  The track length distributions in Fig.~\ref{Distributions} are quite similar.   
A second way to test whether the fall off in the signal/track between the two runs is due to differences in the track length distributions is to use the TallBo simulations.  These  
show that an average of $3.73 \times 10^5$ photons/track fall on the light guides in run~2 and an average $3.76 \times 10^5$ photons/track in run~4, results that incorporate the geometry corrections in Table~\ref{tab:Results}.  
It is unlikely that the drop-off in signal seen from run 2 to run 4 is due differences in the cosmic track samples between the runs.


\subsection{TPB Leaching out of the Light Guides into the LAr}

The last possibility to be considered is that the TPB leached out of the detectors while submerged in LAr, as has been suggested in~\cite{bib:asaadi2019}.  This explanation is unlikely, however, since the response dropped precipitously from run~3 to run~4 but showed almost no change from run~2 to run~3.  If the TPB had leached out of the detectors into the LAr, the drop-off would be expected to be more or less continuous from run~2 to run~4.  

\section{Conclusions}

The analysis of the data from this TallBo experiment clearly shows that there was a drop-off in the light yield seen in both the IU and Fermi technologies after the dewar was emptied of LAr and then refilled.  Neither absorption nor quenching by N$_2$, O$_2$, and H$_2$O contamination can account for the degradation.  Neither the individual Hamamatsu MPPCs nor the passive/active ganging boards appear to have been affected by the thermal cycling.  The path length distributions of the cosmics traversing the dewar appear quite similar in both event samples.

Two possible explanations for this drop-off can be identified.  First, it is possible that the response of the two light guide technologies degraded as the result of thermal cycling.  In this case, most likely the thermal cycling caused physical damage to the light guides that degraded their attenuation lengths, although visual inspection did not expose any obvious crazing.  One impact of this possibility concerns the practice of qualifying light guides for experiments by submerging them in LN2, a practice that may not be optimal for the preservation of performance characteristics of the light guides.

Second, small but differing trace amounts of Xe in the two fills of LAr delivered to the experiment might be responsible for the drop-off.  This explanation could not be evaluated quantitavely since no Xe measurement apparatus was available at TallBo during the experiment.  The impact of this possibility is clear for photon detection in experiments that replenish LAr during the course of the experiment or drain or refill LAr for multiple runs.


\vspace{1.cm}

\noindent {\bf Acknowledgements} 

\noindent This work was supported in part by the Trustees of Indiana University, the DOE Office of High Energy Physics through grant DE-SC0010120 to Indiana University, and grant \#240296 from Broookhaven National Laboratory to Indiana University.  
The authors wish to thank the many people who helped make this work possible. At IU: M.~Gebhard, M.~Lang, J.~Urheim.  At Fermilab: R.~Davis, A.~Hahn, B.~Miner, T.~Nichols, B.~Ramson.  At ANL: G.~Drake.  At MIT: J.~Conrad.  At Eljen Technology: C. Hurlbut. 

This manuscript has been authored by Fermi Research Alliance, LLC under Contract No. DE-AC02-07CH11359 with the U.S. Department of Energy, Office of Science, Office of High Energy Physics. The United States Government retains and the publisher, by accepting the article for publication, acknowledges that the United States Government retains a non-exclusive, paid-up, irrevocable, world-wide license to publish or reproduce the published form of this manuscript, or allow others to do so, for United States Government purposes.

\hfil

\bibliographystyle{elsarticle-num}

\bibliography{ThermalCycling}

\begin{thebibliography}{10}
\expandafter\ifx\csname url\endcsname\relax
  \def\url#1{\texttt{#1}}\fi
\expandafter\ifx\csname urlprefix\endcsname\relax\def\urlprefix{URL }\fi
\expandafter\ifx\csname href\endcsname\relax
  \def\href#1#2{#2} \def\path#1{#1}\fi

\bibitem{bib:DUNE-CDR-vol1}
R.~Acciarri, et~al., Long-{B}aseline {N}eutrino {F}acility ({LBNF}) and {D}eep
  {U}nderground {N}eutrino {E}xperiment ({DUNE}) {C}onceptual {D}esign {R}eport
  vol.1: {T}he {LBNF} and {DUNE} {P}rojects, arXiv:1601.05471.

\bibitem{bib:howard}
B.~Howard, et~al., A novel use of light guides and wavelength shifting plates
  for the detection of scintillation photons in large liquid argon detectors,
  NIM A907 (2018) 9.

\bibitem{bib:MITbars}
L.~Bugel, et~al., Demonstration of a lightguide detector for liquid argon
  {TPC}s, NIM A640 (2011) 69.

\bibitem{bib:MITbars2}
Z.~Moss, et~al., Improved {TPB}-coated light guides for liquid argon {TPC}
  light detection systems, JINST 10 (2015) P08017.

\bibitem{bib:MITN2}
B.~Jones, et~al., A measurement of the absorption of liquid argon scintillation
  light by dissolved nitrogen at the part-per-million level, JINST 8 (2013)
  P07011.

\bibitem{bib:O2Contamination}
R.~Acciarri, et~al., Oxygen contamination in liquid argon:combined effects on
  ionization electron charge and scintillation light, JINST 5 (2010) P05003.

\bibitem{bib:O2contamCross}
R.~Goldstein, F.~Mastrup, Absorption coefficients of the {O}2
  {S}chumann--{R}unge continuum from 1270 {{\AA}} - 1745 {{\AA}} using a new
  continuum source, JOSA 56 (1966) 765.

\bibitem{bib:H2O}
K.~Mavrokoridis, et~al., Argon purification studies and a novel liquid argon
  re-circulation system, JINST 6  P08003.

\bibitem{bib:H2OcontamCross}
K.~Watanabe, M.~Zelikoff, Absorption coefficients of water vapor in the vacuum
  ultraviolet, JOSA 43 (1953) 9.

\bibitem{bib:TallBo}
D.~Whittington, S.~Mufson, B.~Howard, Scintillation light from cosmic-ray muons
  in liquid argon, JINST 11 (2016) P05016.

\bibitem{bib:neumeierXe}
A.~Neumeier, et~al., Attenuation of vacuum ultraviolet light in pure and
  xenon-doped liquid argon: An approach to an assignment of the near-infrared
  emission from the mixture, EPL 111 (2015) 12001.

\bibitem{bib:N2Contamination}
R.~Acciarri, et~al., Effects of nitrogen contamination in liquid argon, JINST 5
  (2010) P06003.

\bibitem{bib:kubota}
S.~Kubota, et~al., Liquid and solid argon, krypton and xenon scintillators, NIM
  196 (1982) 101.

\bibitem{bib:scintYield2}
T.~Doke, et~al., Absolute scintillation yields in liquid argon and xenon for
  various particles, Jpn.J.Appl.Phys. 41 (2002) 1538.

\bibitem{bib:PDG}
C.~Patrignani, et~al., The {R}eview of {P}article {P}hysics, Chin. Phys. C 40
  (2017) 100001.

\bibitem{bib:Wahl}
C.~Wahl, et~al.,
  \href{http://stacks.iop.org/1748-0221/9/i=06/a=P06013}{Pulse-shape
  discrimination and energy resolution of a liquid-argon scintillator with
  xenon doping}, JINST 9 (2014) P06013.
\newline\urlprefix\url{http://stacks.iop.org/1748-0221/9/i=06/a=P06013}

\bibitem{bib:LArScint}
T.~Heindl, et~al., The scintillation ofl iquid argon, EPL 91 (2010) 62002.

\bibitem{bib:Belov}
A.~Belov, et~al., Luminescence of oxygen-rare gas exciplex compounds in rare
  gas matrices, J.Lumin 91 (2000) 107.

\bibitem{bib:Johnson}
D.~E. Johnson, Rare gas {-} oxygen emission bands and rare gas continua in the
  {UV} and {VUV}, Chem.Phys.Lett. 238 (1995) 71.

\bibitem{bib:morozov}
A.~Morozov, et~al., Ultraviolet emission from argon water-vapor mixtures
  excited with low-energy electron beams, APL 86 (2005) 011502.

\bibitem{bib:asaadi2019}
J.~Asaadi, et~al., Emanation and bulk fluorescence in liquid argon from
  tetraphenyl butadiene wavelength shifting coatings, JINST 14 (2019) P02021.

\end{thebibliography}

\newpage

\appendix
\section{The TallBo Experiment}
\label{sec:TallBoExperimentAppendix}

\begin{figure}[h]
\centering
\includegraphics[width=.6\textwidth]{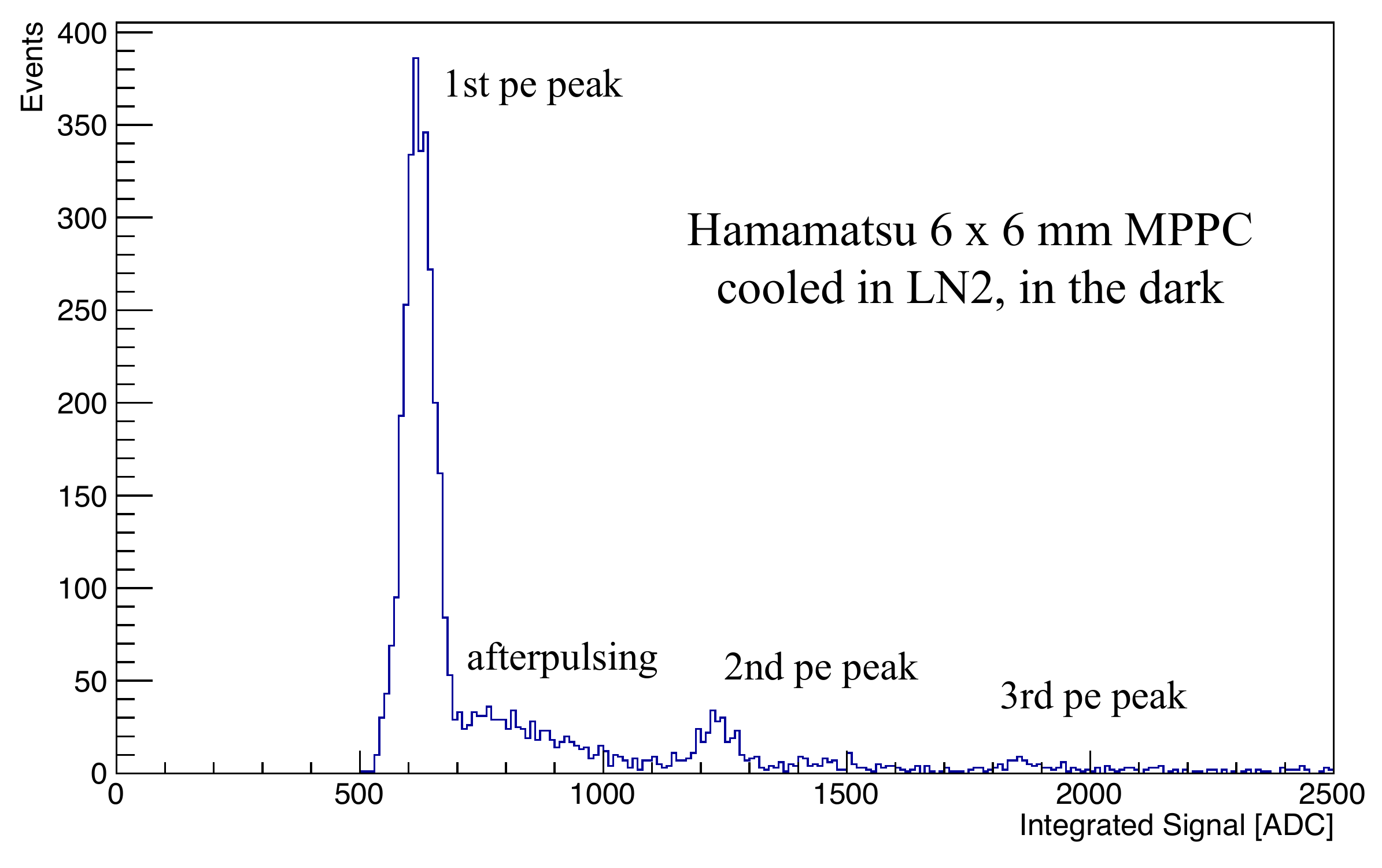}
\caption{The dark spectrum of a previously uncooled MPPC in LN2 that has been biased at 44.5~V as measured in the lab.  The single p.e. peaks are marked.  Also marked is the signal due to afterpulsing.}
\label{testMPPCAppendix}
\end{figure}

Fig.~\ref{testMPPCAppendix} shows the 
dark spectrum of a previously uncooled MPPC in LN2 that has been biased at 44.5~V.  The first, second, and third photoelectron (p.e.) peaks are clearly visible.  Also labeled is the signal from afterpulsing.  Afterpulsing is a second avalanche in the same pixel as the primary avalanche.  They are smaller than a standard avalanche because they occur before the cell can fully recover.  Afterpulsing can seriously blur primary event signals when it becomes significant.

\begin{figure}[h] 
\centering
\includegraphics[width=.45\textwidth]{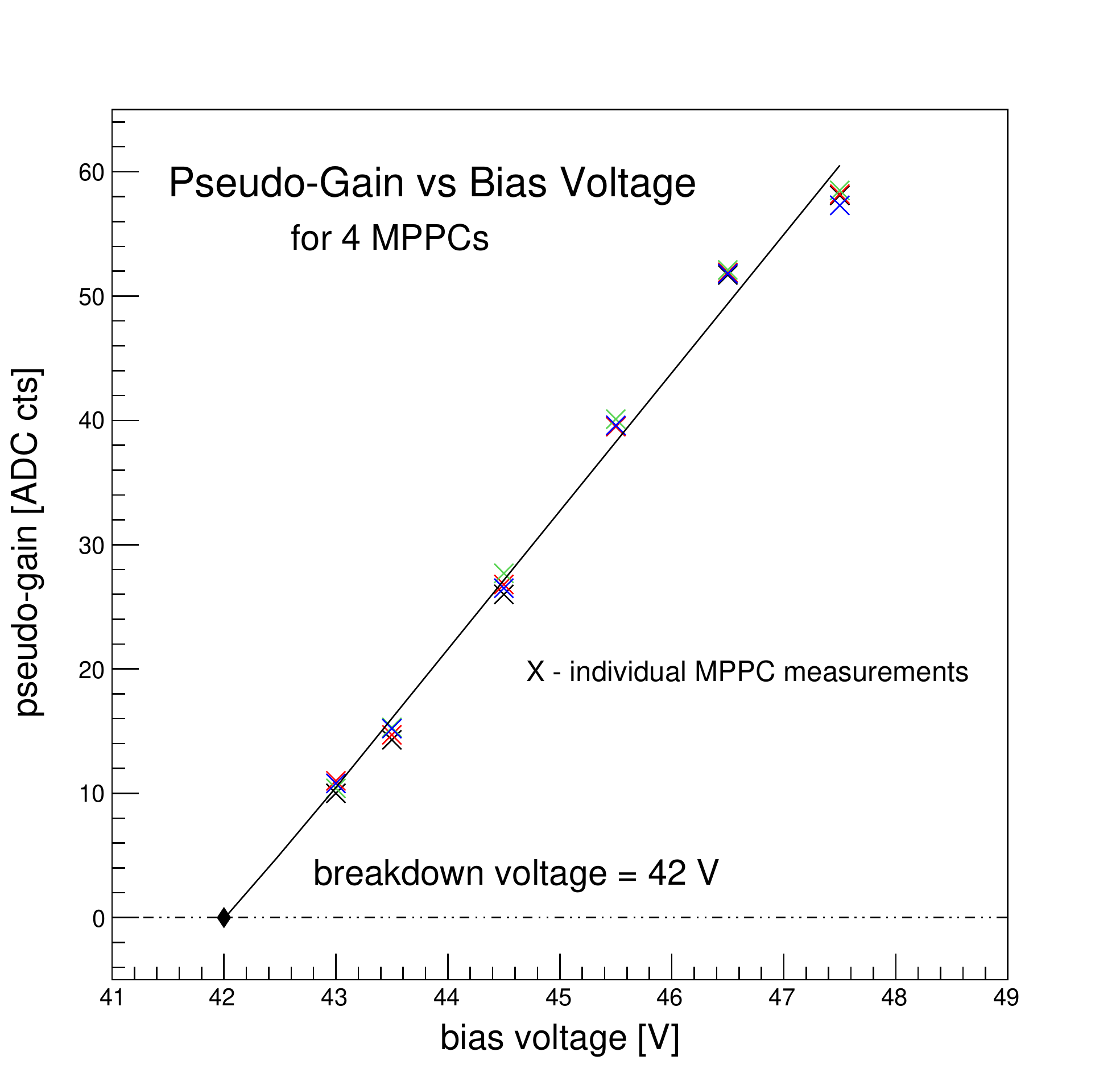}
\caption{The pseudo-gain, or the difference in the number of ADC counts between the first and second p.e. peaks,  as a function of bias voltage for 4 previously uncooled MPPCs.  The x~intercept of a least-squares straight line fit to the 4 pseudo-gain measurements determines the breakdown voltage.}
\label{breakdownVoltageAppendix}
\end{figure}

The breakdown voltage for the MPPCs is 42~V, determined by plotting the ``pseudo-gain'', as a function of bias voltages ranging from 43V - 47.5V, as shown in Fig.~\ref{breakdownVoltageAppendix} for 4 previously 
uncoold MPPCs.  The pseudo-gain is the difference in the number of ADC counts between the first and second p.e. peaks, a quantity that is related to the MPPC gain by a multiplicitive constant.  The breakdown voltage was determined as the x~intercept of a least-squares straight line fit to the mean of the pseudo-gain measurements for the 4 MPPCs as a function of bias voltage. 

\begin{figure}[t!]
\centering
\includegraphics[width=2.3in,height = 1.9in]{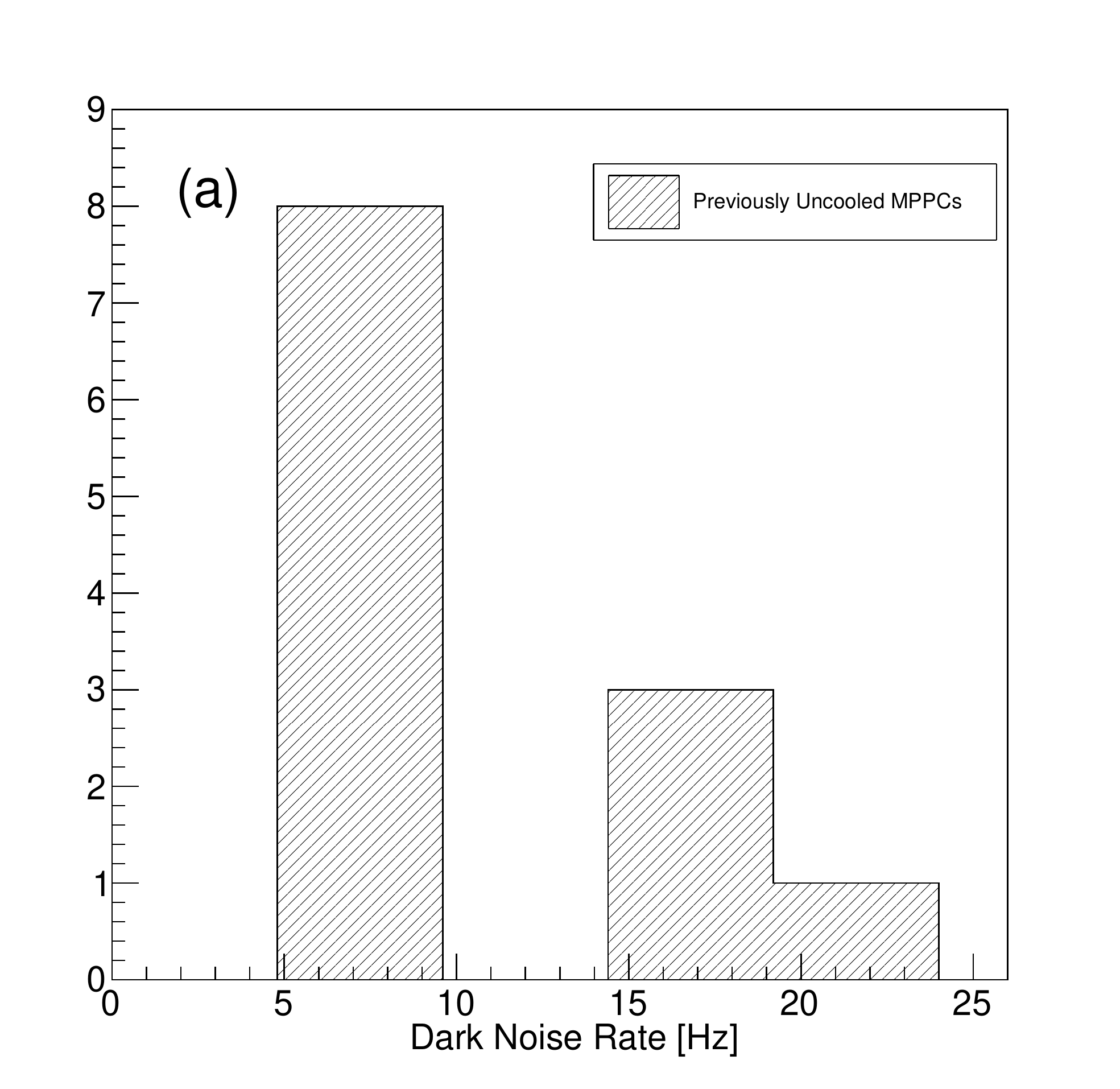}
\includegraphics[width=2.3in,height = 1.9in]{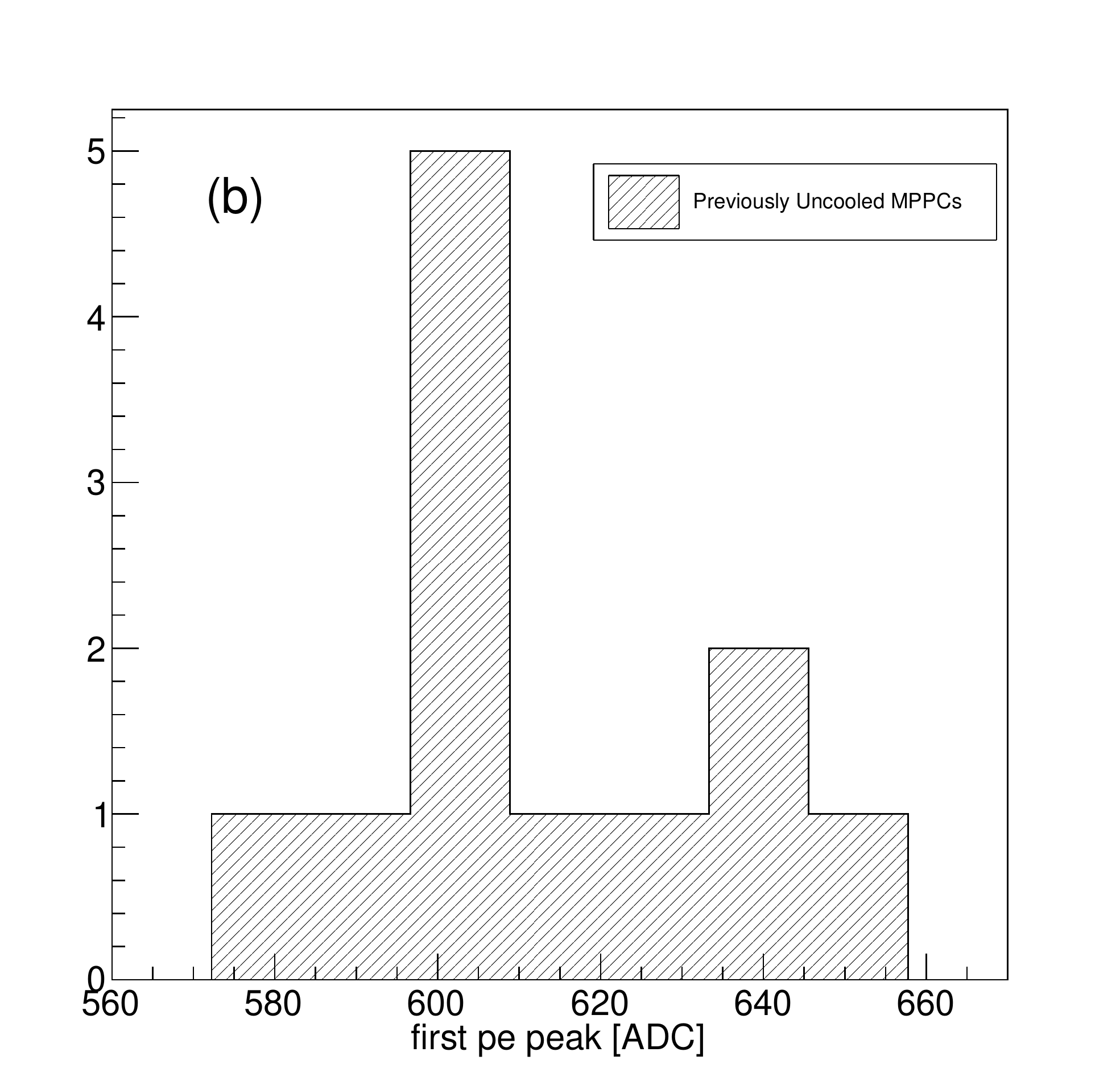}\\
\includegraphics[width=2.3in,height = 1.9in]{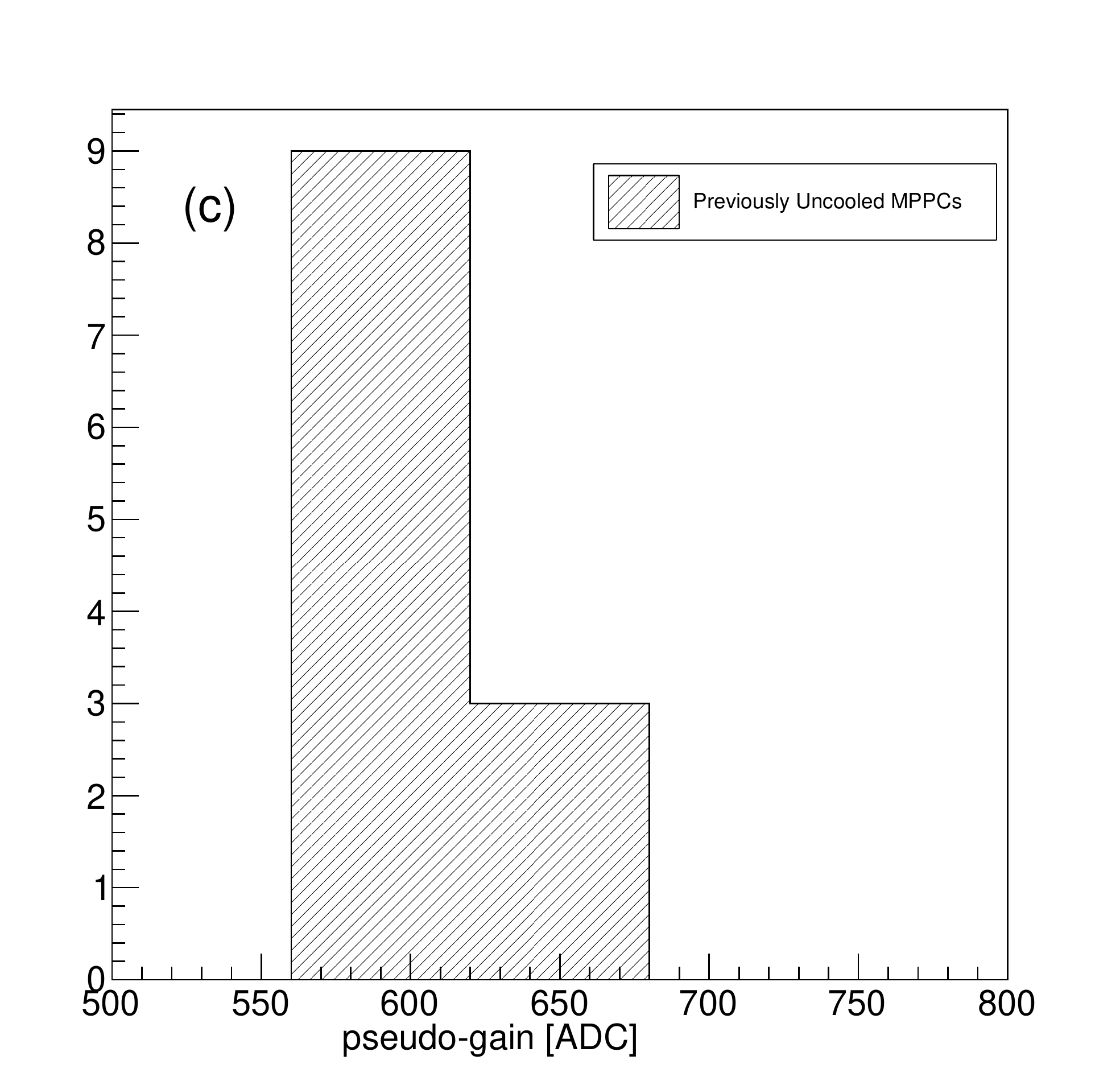}
\includegraphics[width=2.3in,height = 1.9in]{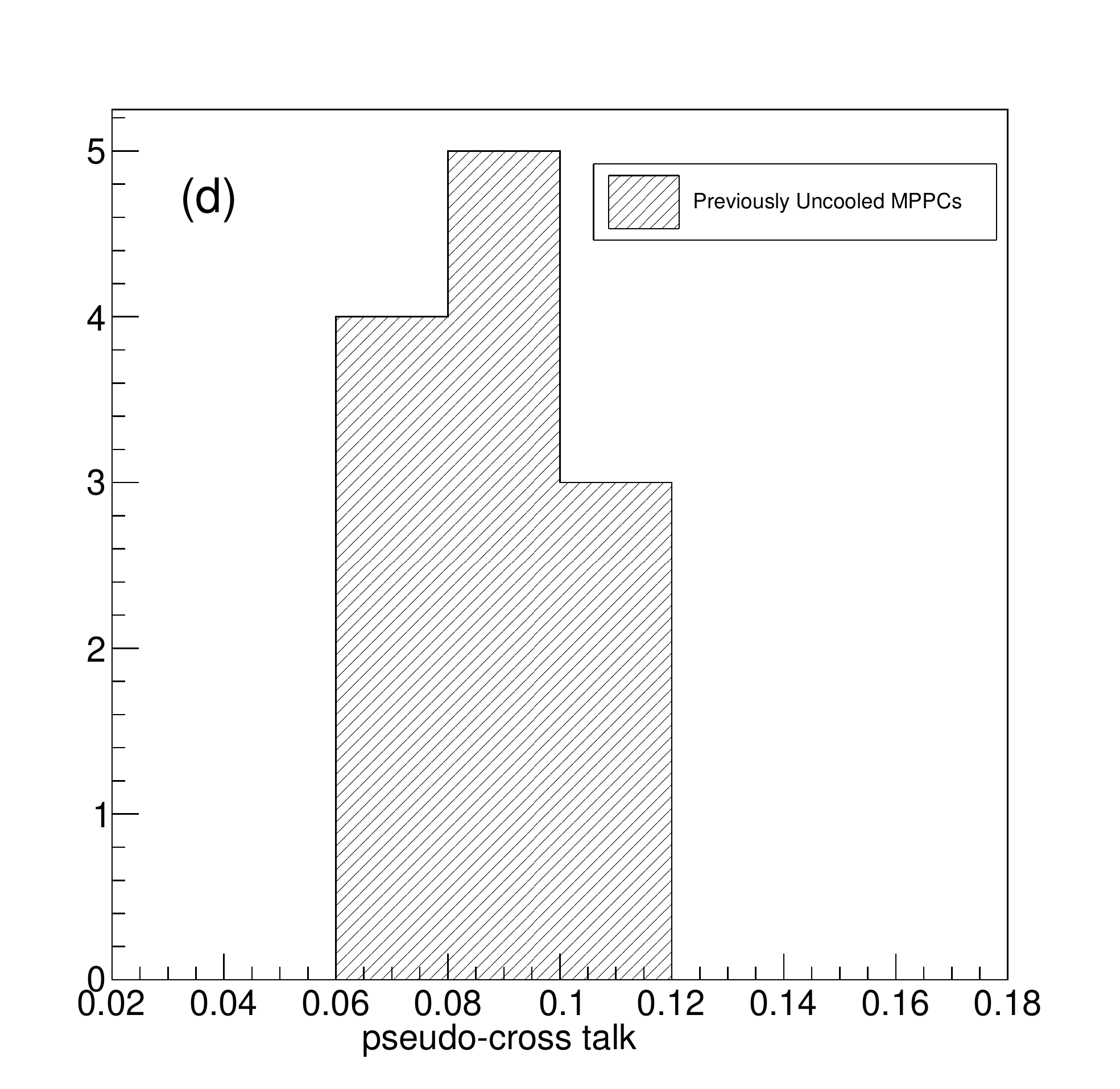}\\
\caption{Laboratory measurements of the the performance characteristics of 12 previously uncooled Hamamatsu MPPCs immersed in LN2 biased at 44.5~V.  (a) The dark noise, characterized as the sum of the ADC counts divided by the data acquisition time; (b) the number of ADC counts at the peak of the first p.e. peak; (c) the pseudo-gain or the number of ADC counts between the first and second p.e. peaks; and (d) the pseudo-cross talk determined as the ratio of the number of ADC counts in the peak of the second p.e. peak to the number of ADC counts in the peak of the first p.e. peak.}
\label{MPPCcharacteristicsAppendix}
\end{figure}

Laboratory measurements of the the performance characteristics of 12 previously uncooled Hamamatsu MPPCs immersed in LN2 are shown in Fig.~\ref{MPPCcharacteristicsAppendix}.  The MPPCs were biased at 44.5~V.  
These measured characteristics are with reference to  Fig.~\ref{testMPPCAppendix}.   The dark noise (a) was computed as the sum of the ADC counts in the dark spectrum divided by the data acquisition time.  The MPPCs run quietly in LN2 with rates in the range of 5-25 Hz, compared with their MHz dark rates at room temperature.  The first p.e. peak (b) is the number of ADC counts at the peak of the first p.e. peak.  This calibration of the individual MPPCs showed variations of approximately 15$\%$.  The ``pseudo-gain'' (c) is the difference in the number of ADC counts between the first and second p.e. peaks, a quantity that is related to the gain by a multiplicitive constant.  
Cross talk events occur when a photon emitted during the electron avalanche in one pixel is re-absorbed by another pixel elsewhere on the SiPM and induces a second avalanche in immediate coincidence with the first.  The ``pseudo-cross talk'' (d) is the ratio of the number of ADC counts in the peak of the second p.e. peak to the number of ADC counts in the peak of the first p.e. peak and estimates the contribution of cross-talk to the event signal.  It is typically in the 10$\%$ range.

\begin{figure}[h]
\centering
\includegraphics[width=2.2in,height = 2in]{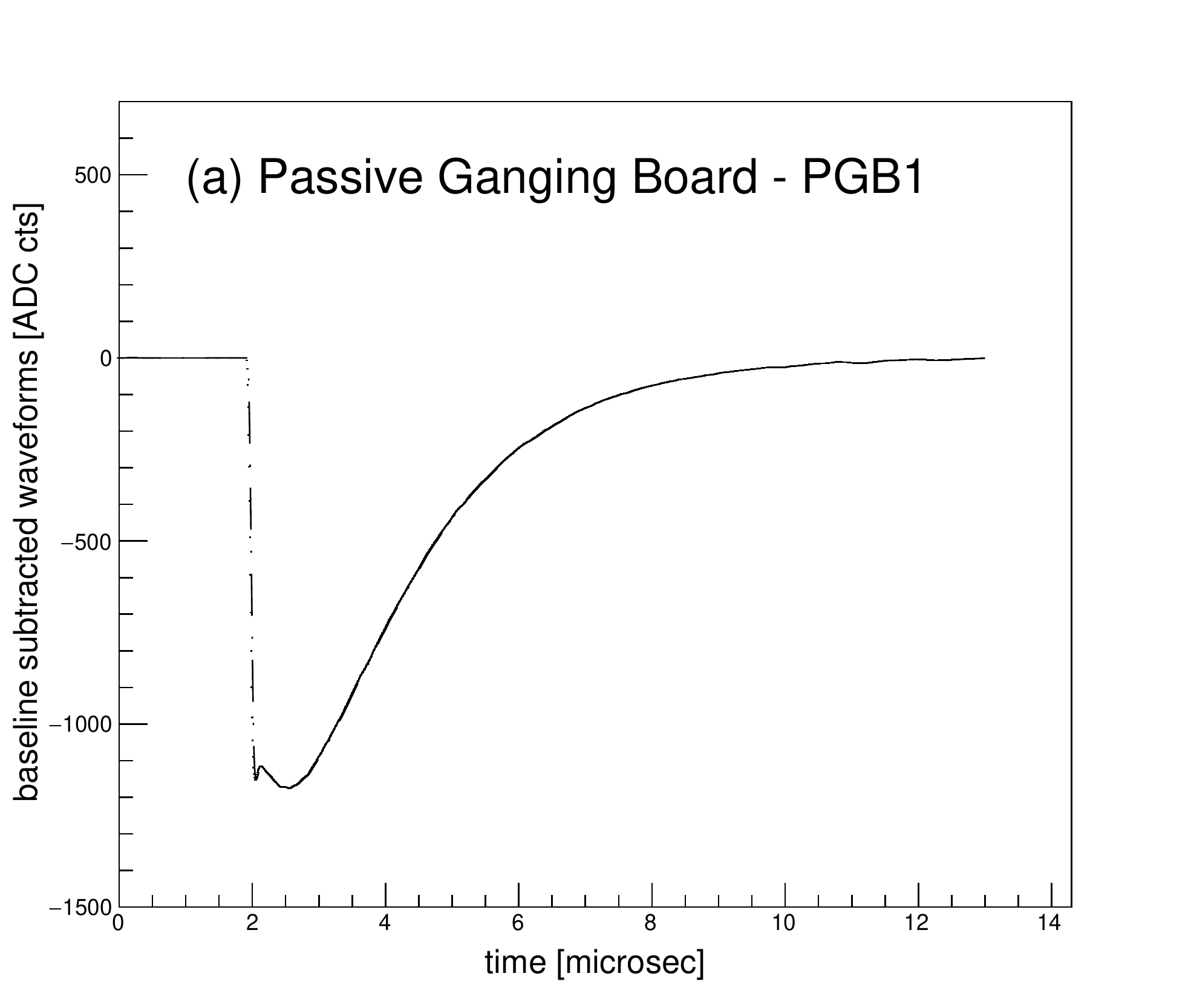}
\includegraphics[width=2.2in,height = 2in]{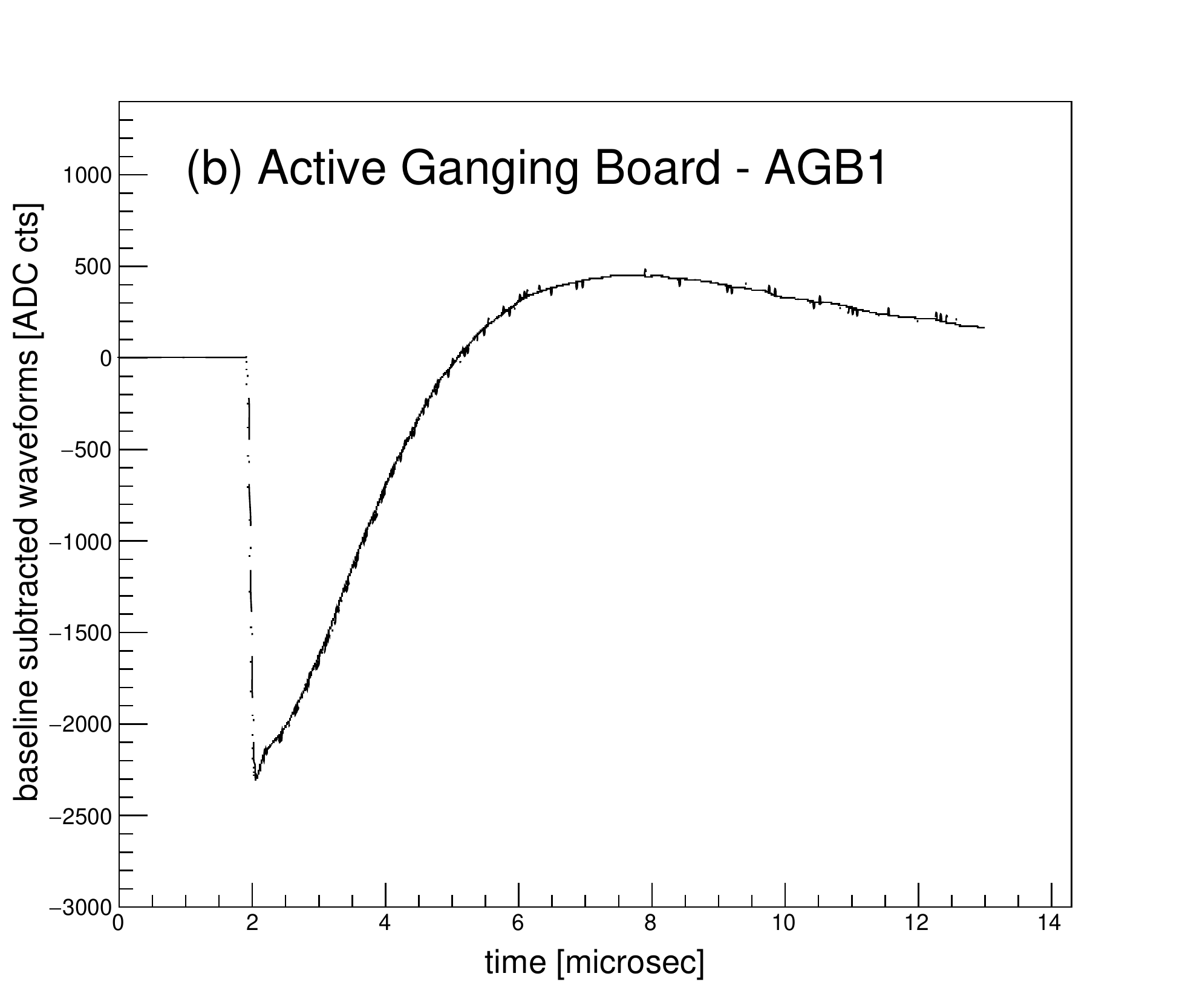}
\caption{The baseline subtracted mean of 25 waveforms from typical single track muon events crossing TallBo and which pass all analysis cuts as read out by (a) passive ganging board PGB1 and (b) active ganging board AGB1 (Fig.~\ref{exptLayout}) during run 2.  In these waveforms each bin is 6.67 ns wide and the waveforms shown are $\sim$10.7~$\mu$s in length.  The waveforms from the readout of the Fermilab technology are qualitatively very similar.}
\label{boardWFsAppendix}
\end{figure}

The passive ganging boards, PGB1 and PGB2 in Fig.~\ref{exptLayout}, connect the 12 MPPCs in parallel.  
Panel (a) in Fig.~\ref{boardWFsAppendix} shows the baseline subtracted mean of 25 waveforms from typical single track muon events crossing TallBo as read out by the passive board PGB1 during run 2. The waveforms from the readout board PGB2 on the Fermilab technology are qualitatively very similar.  The increased capacitance is seen in the long exponential tail on the waveform of $\approx$ 6.5 $\mu$s.  Single MPPCs have an exponential tail more typically $\approx$ 500~ns.  The artifact at the waveform minimum is thought to be the result of mismatches in impedance at the board-SSP coupling. 

\begin{figure}[h]
\centering
\includegraphics[width=2.5in,height = 1.3in]{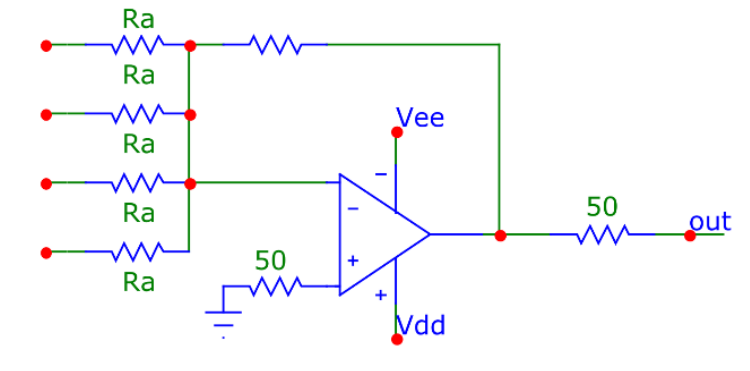}\\
\includegraphics[width=4.6in,height = 3.4in]{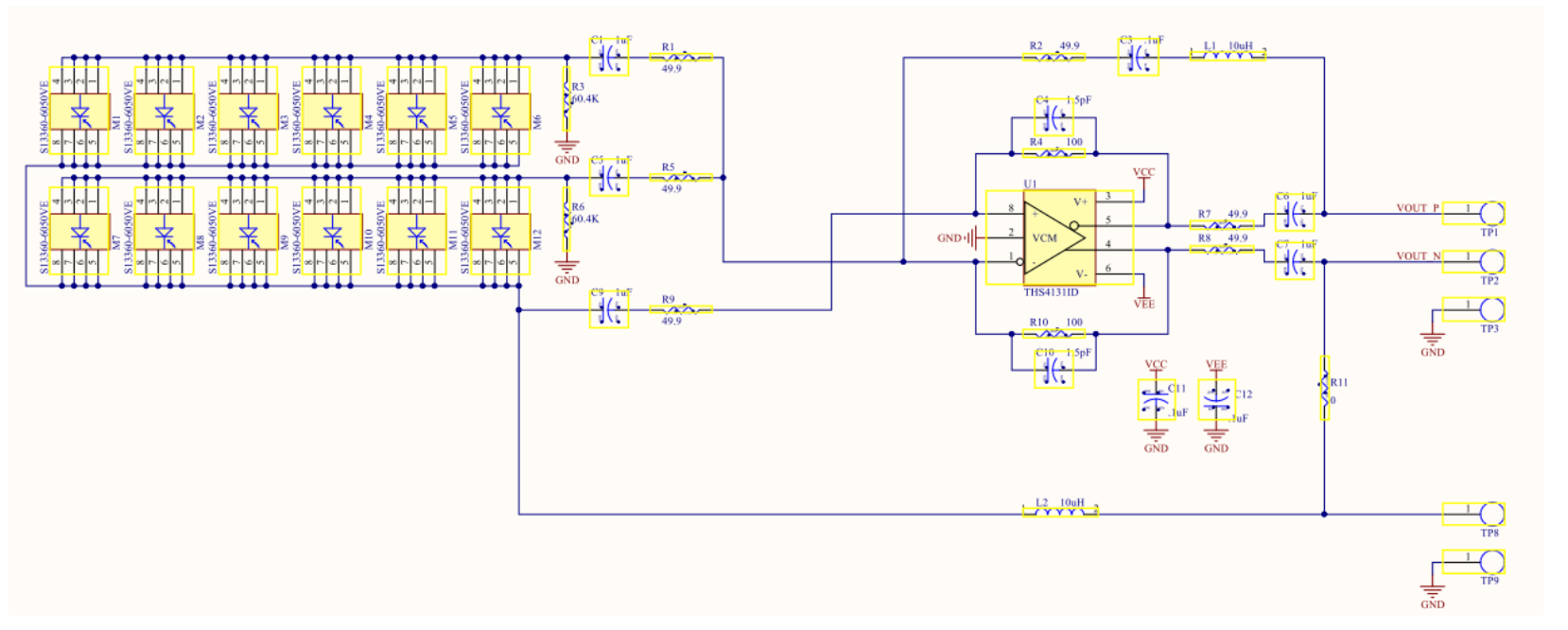}
\caption{{\it top}: The active summing node used in the design of the active ganging board; {\it bottom}: the active ganging board.}
\label{activeBoardDesignAppendix}
\end{figure}

The active summing node for the active ganging board shown in the {\it top} panel of 
Fig.~\ref{activeBoardDesignAppendix}.  
In this circuit the negative input of the OpAmp provides a virtual ground that decouples the signals coming on each input.  The gain of the amplifier can then be adjusted by the ratio between the feedback resistor and Ra.  
There are several challenges to designing this active summing node: it must work in the cold (87K); it must not boil the LAr by dissipating too much heat; it must be high bandwidth to keep the MPPC signal integrity; and it must not contribute significantly to the total noise budget.  The circuit used in this experiment employed a Texas Instruments THS4131ID\footnote[3]{https://www.ti.com/store/ti/en/p/product/?p=THS4131ID} OpAmp which has 150 MHz of bandwidth for unitary gain and 1.3~nV/$\sqrt{{\rm (Hz)}}$ of noise.  In this circuit the Op Amp summed 2 rows of 6 MPPCs each.  
The active summing board is shown in the {\it bottom} panel of Fig.~\ref{activeBoardDesignAppendix}.

Two hodoscope modules were installed on opposite sides of the TallBo dewar to select single-track cosmic-ray muons passing through the LAr volume~\cite{bib:TallBo}.  Each hodoscope module consists of 64 2-inch diameter barium-fluoride crystals, coated with TPB and arranged in an 8$\times$8 array. Each crystal is monitored by a 3$''$ PMT. 
Since the hodoscope modules were originally designed to detect bremsstrahlung photons in the CREST balloon flight experiment, they are very sensitive to extraneous photon activity around our experiment. To reject this $\gamma$~ray activity, a pair of plastic scintillator panels covering the entire face of a hodoscope module were placed between each hodoscope module and the TallBo dewar. That is, there were 4 total scintillator panels, two on each side of the TallBo dewar. These 4 panels were individually read out by a PMT salvaged from the QuarkNet program at FNAL. The SSP readout was triggered by four-fold coincidence logic that required at least one hit in each hodoscope module on opposite sides of the dewar, as well as one hit in the adjacent scintillator plane. Typically this trigger indicates that each event contains at least one charged particle passing through the LAr.  Events were further filtered offline to reject showers by requiring one and only one hit on a PMT in each hodoscope module. Single-track events crossing from one side of the frame to the other were rejected in order to exclude any tracks that could pass through a light guide.

\section{Operations}

The first two runs differed only by cable switches.  
There were actually two monitor photosensors on the frame of two different types: the Hamamatsu MPPC shown in Fig.~\ref{exptLayout} and a SensL SiPM like those used in our previous  experiments at PAB~\cite{bib:howard,bib:TallBo}.  During data analysis, it was concluded that the SensL SiPM did not contribute useful information but this was not established during the experiment.  At the outset of data-taking, one of the channels assigned to the monitor SiPMs failed. 
Between run 1 and run 2 the readout channel for the two different monitor SiPMs were exchanged, which resulted in a cable switch.  Both of these runs read out front side tracks.  

\subsection{Data Analysis}
\label{AnalysisAppendix}

\begin{figure}[h]
\centering
\includegraphics[width=2.4in,height = 2.in]{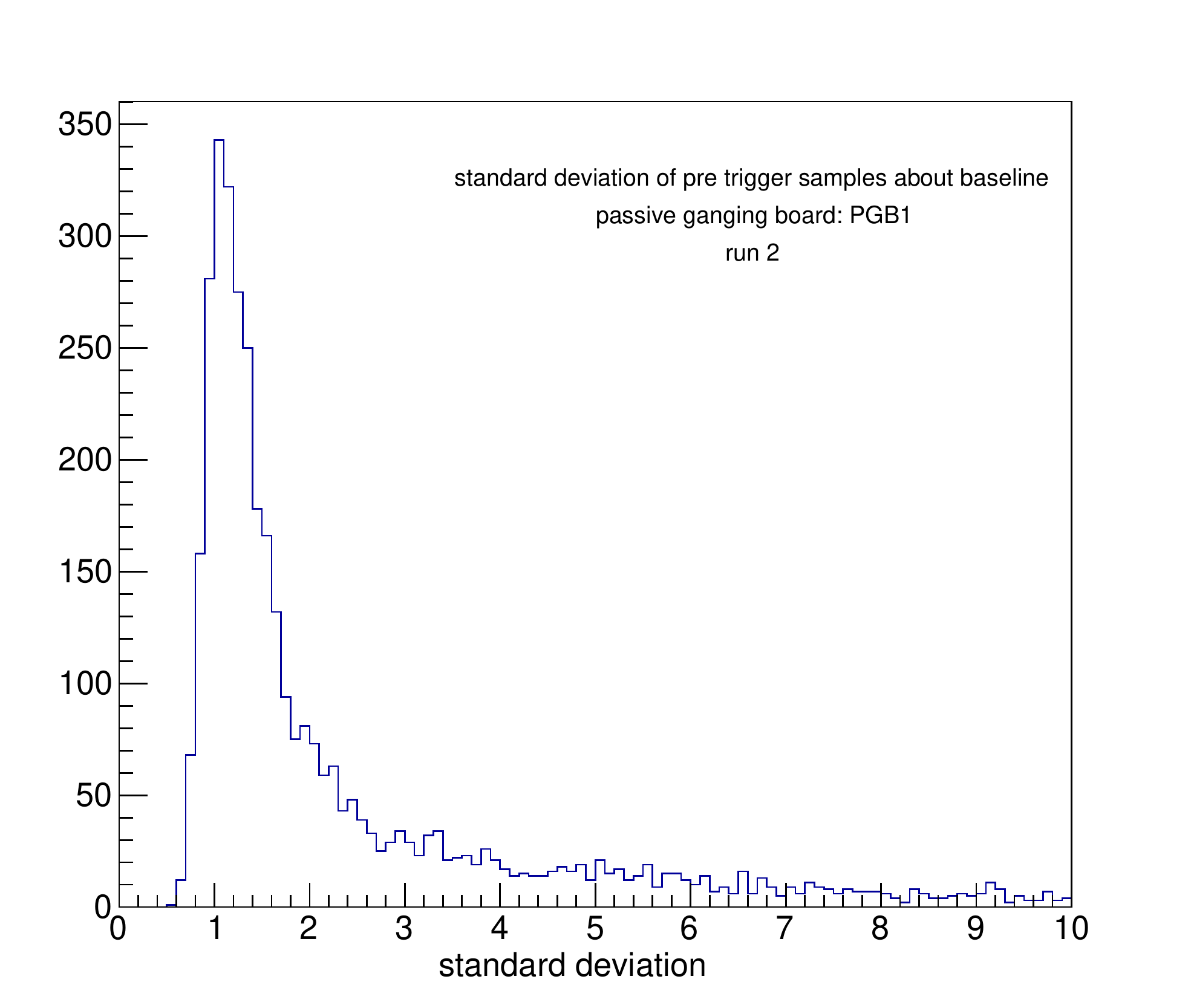}
\includegraphics[width=2.4in,height = 2.in]{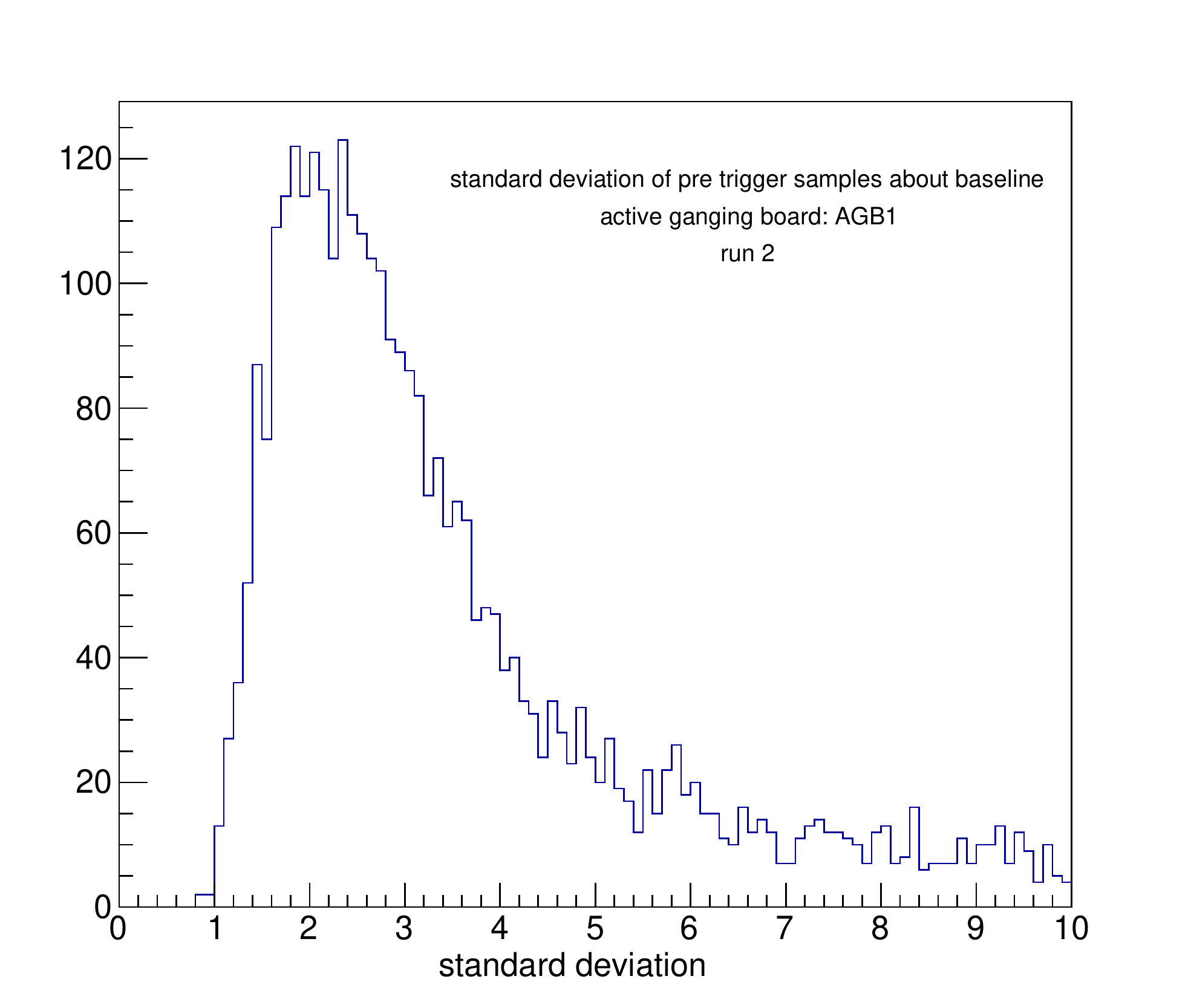}
\caption{The distributions of the standard deviation of the pre trigger samples about the baseline for the passive readout board PGB1 and the active board AGB1 from run 2.  }
\label{SDrun2Appendix}
\end{figure}

The data analysis began by filtering the waveforms of the selected events with an 11-point running mean.  This procedure smooths out fluctuations along the waveform that can result from the decreased signal to noise due to ganging.  Fig.~\ref{boardWFsAppendix} in the Appendix shows the readout retains $\sim$2 $\mu$s of data before the trigger and the mean of the pre-trigger data was used to calculate the waveform baseline.  A cut was then applied on the standard deviation of the pre trigger samples about the baseline to remove anomalous waveforms with spurious shapes or unusually large fluctuations.  These anomalous waveforms were found to correlate strongly with large values of the standard deviation.  Fig.~\ref{SDrun2Appendix} shows the standard deviations of all waveforms from boards PGB1 and AGB1 from run 2.  
For both distributions, the cut was made at a standard deviation of 2, which was found to eliminate the vast majority of anomalous waveforms.  PGB2 and AGB2 have similar distributions and the cut was also made at a standard deviation of 2.  Qualitatively similar histograms were found for the readout boards in runs 3 and 4, and the cut was again made at a standard deviation of 2 for consistency. 

One characteristic of the histograms in Fig.~\ref{SDrun2Appendix} is apparent.  Comparing the two distributions, a larger fraction of the waveforms from AGB1 have large standard deviations.  These differences are also seen when comparing PGB2 and AGB2.  In runs 3 an 4, the differences are often even more striking.  This suggests that the active boards are often shaping and distorting the waveforms in significant ways, which is likely the result of amplifier noise.  It also means that the standard deviation cut removes a far larger fraction of waveforms read out by the active boards than from the passive boards.

The choice to integrate the waveforms from AGB1 and AGB2 only out to the overshoot reflects the fact that the overshoot was found to occur at a position (time) along the waveform that is approximately independent of signal strength, suggesting that the overshoot is an artifact of how the active ganging board shapes the waveform and the piece of the waveform after the overshoot does not add to the charge information in the waveform.  A lab test of this hypothesis was made at IU after the experiment was completed.   In these tests the boards were immersed in LN2 and then flashed with LED pulses of variable width.  The assumption was made that the number of photons falling on the boards was directly proportional to the LED pulse width.  The range in pulse widths reflect the transit times of a muon assuming the light is dominated by early light~\cite{bib:TallBo}.  The waveforms from the passive boards were integrated out to 10.7 $\mu$s.  The waveforms from the active boards were integrated out to the stable overshoot point.  Since all charge information is contained in the integral of the passive waveforms, it should be linear with LED pulse width.  If the charge information in the integrated waveforms from the active boards is a constant fraction of the total contained in the integrated wavefrorms from the passive boards, it should also be linear with LED pulse width and have the same slope. 
\begin{figure}[h]
\centering
\includegraphics[width=2.6in,height = 2.4in]{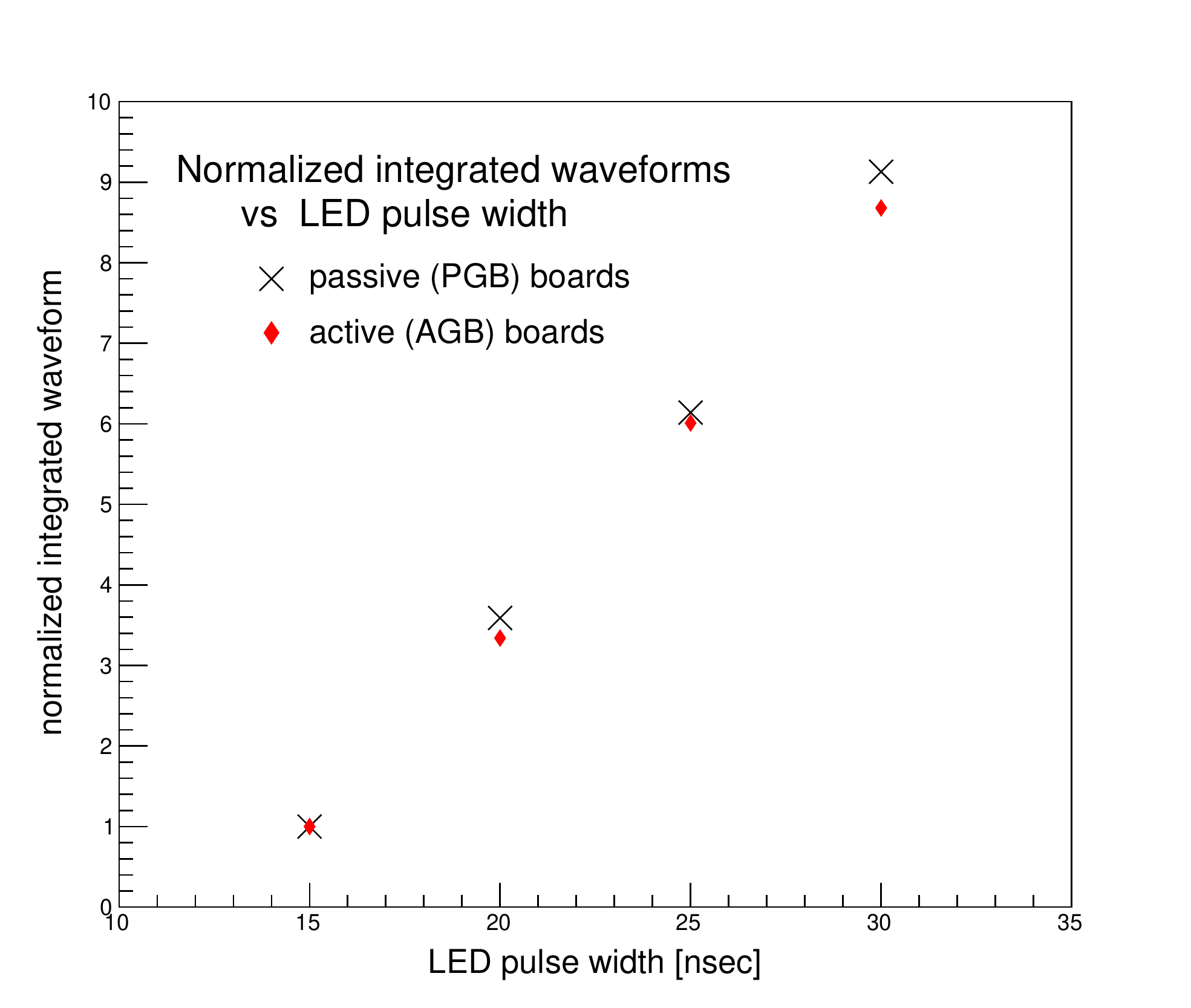}
\caption{The comparison of the integrated waveforms from passive and active readout boards from the TallBo experiment.  The boards were tested in LN2 in the lab by flashing them with an LED having different pulse widths.  The results obtained from the two passive and the two active boards were averaged.  The mean integrated waveforms were normalized to their values are 15 ns.}
\label{activePassiveAppendix}
\end{figure}
Fig.~\ref{activePassiveAppendix} shows this is indeed the case.  It was not feasible to determine whether all charge information is contained in the integral out to the overshoot.  The test dewar was small and reflections off the stainless steel sides made the results dependent on the exact placement of the boards in the dewar.  The exact position of the boards in the dewar from test to test proved impossible to reproduce exactly.  For this reason, the comparison of the passive boards and the active boards was made with the means of the boards and the results were normalized to their values at 15 ns.  Nevertheless, Fig.~\ref{activePassiveAppendix} does make the case that the integrated waveforms from passive boards and the active boards out to the overshoot yields consistent information about the response of the detectors. 

\begin{figure}[h]
\centering
\includegraphics[width=2.5in,height = 2.1in]{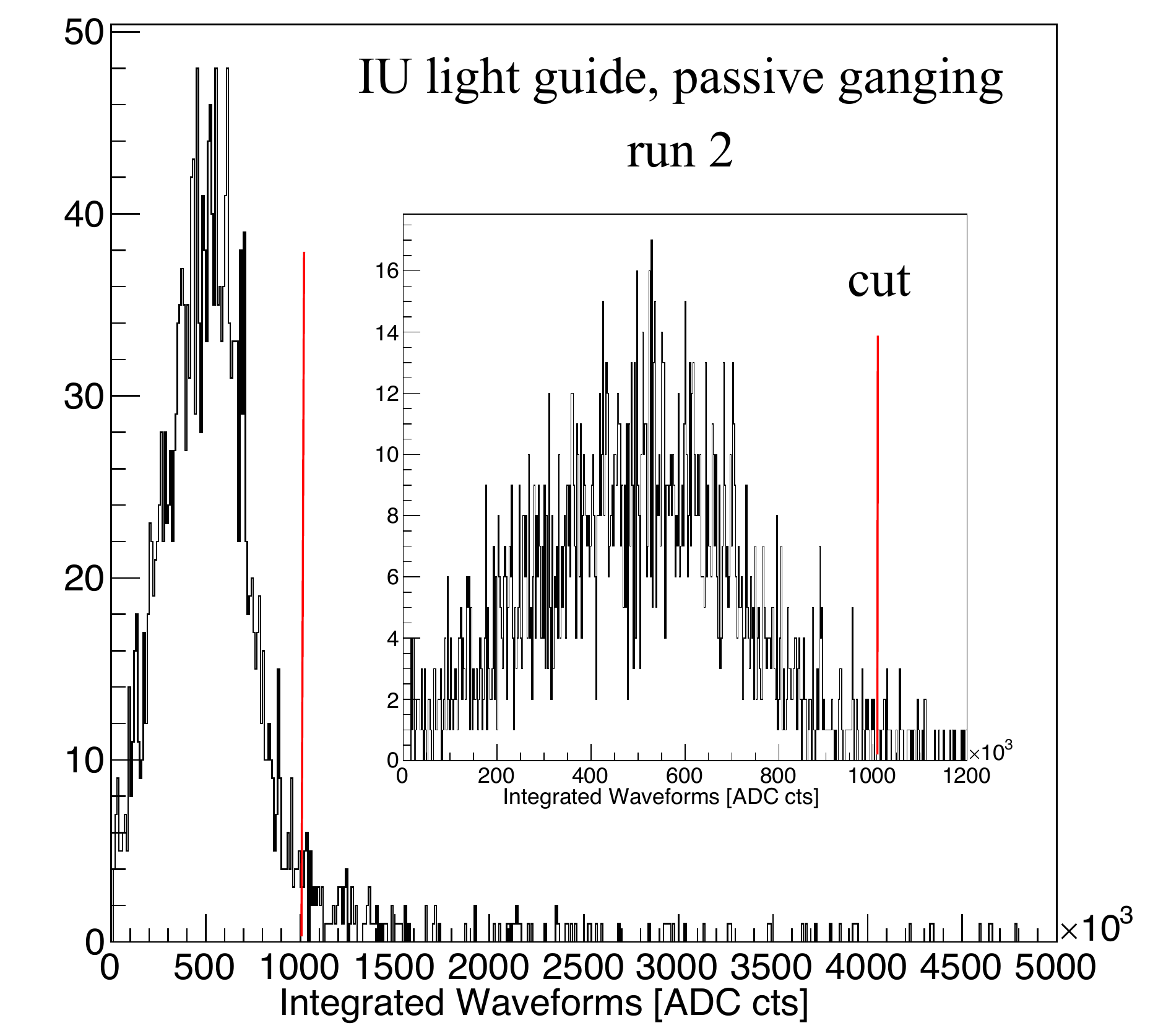}
\includegraphics[width=2.5in,height = 2.1in]{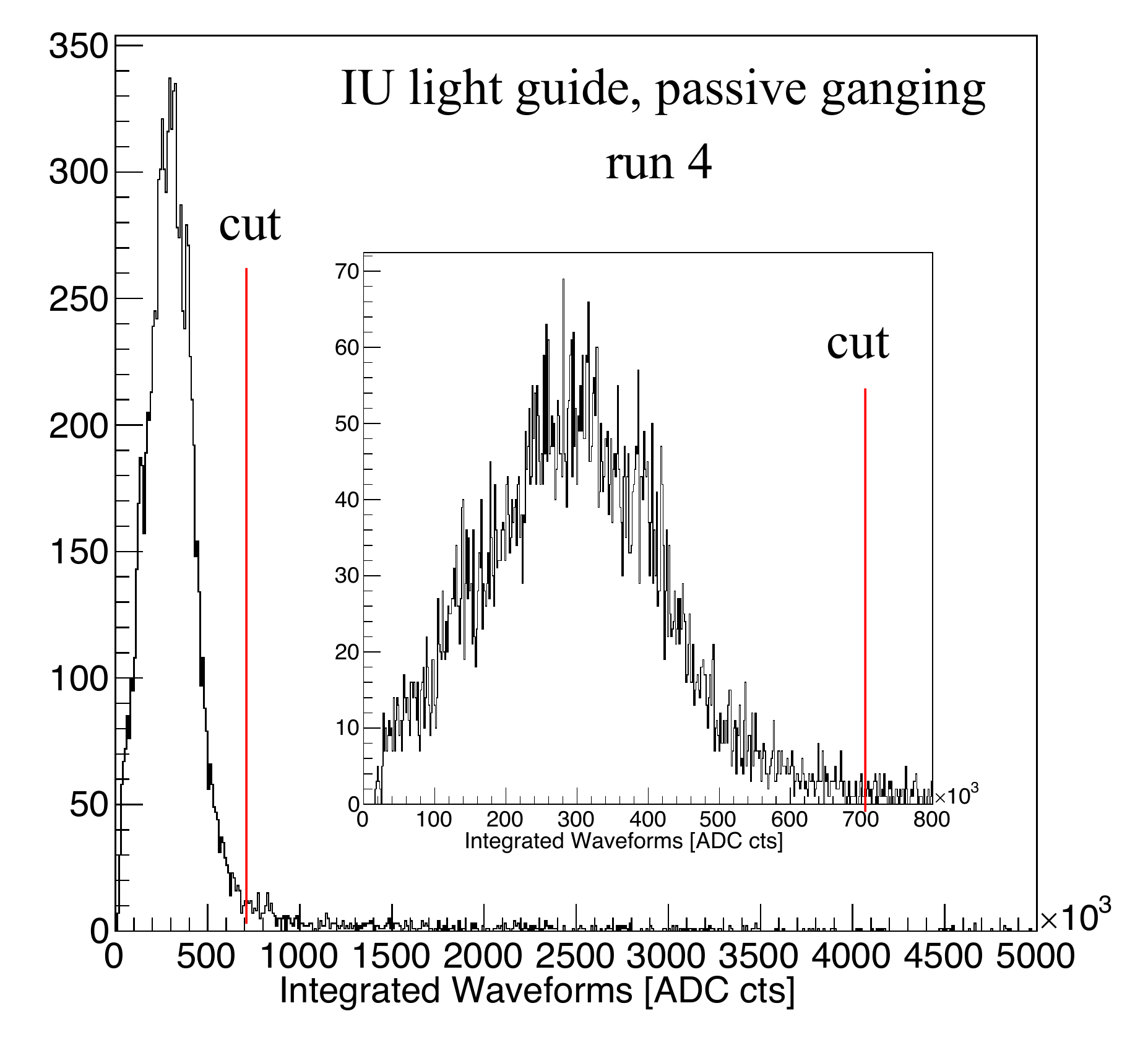}
\caption{The integrated waveforms from the passive boards on the IU light guide in runs 2 and 4.  The long high tail in the integrated are apparent.  The muon ``bump'' for each run is shown in the inset.  The cut on the tail is marked.}
\label{IUrun2-4Appendix}
\end{figure}

The integrated waveforms from the passive and active boards on the IU light guide and the Fermi light guide were put into separate histograms for runs 2, 3, and 4.  Examples of the histograms for the integrated waveforms 
for the passive boards on the IU light guide in runs 2 and 4 are shown in Fig.~\ref{IUrun2-4Appendix}.  

\section{Discussion}

\subsection{Contamination}

There is evidence of additional components of scintillation light from O$_2$ contamination at 200~nm and 577~nm~\cite{bib:LArScint,bib:Belov,bib:Johnson}. 

For the component at 577~nm~\cite{bib:Belov,bib:Johnson}, there is no information on absorption/emission by TPB from 577~nm photons.  It is likely to be small.  The Eljen EJ-280 light guides\footnote[4]{http://www.eljentechnology.com} used in the IU technology do not appear to absorb at 577 nm.  For both the IU technology and the Fermi technology, any 577~nm light that might be absorbed would be waveshifted to an even longer wavelength where the MPPC response drops significantly.  The component at 577~nm is unlikely to play a role in this analysis.

For the component at 200~nm, \cite{bib:LArScint} does not give the amount of impurities in their argon and further suggests the 200~nm component makes up only a small fraction of the scintillation yield.  Since the relative scintillation yield between 128~nm and 200~nm depends on the technology used to measure it, it is hard to assess just how much scintillation light this component yields.  But the fact that scintillation light detected does not rise appreciably from run 2 to run 3 when the O2 contamination increases by an order of magnitude strongly suggests it is not significant.

\subsection{MPPC Response}

\begin{figure}[h]
\centering
\includegraphics[width=2.3in,height = 2.13in]{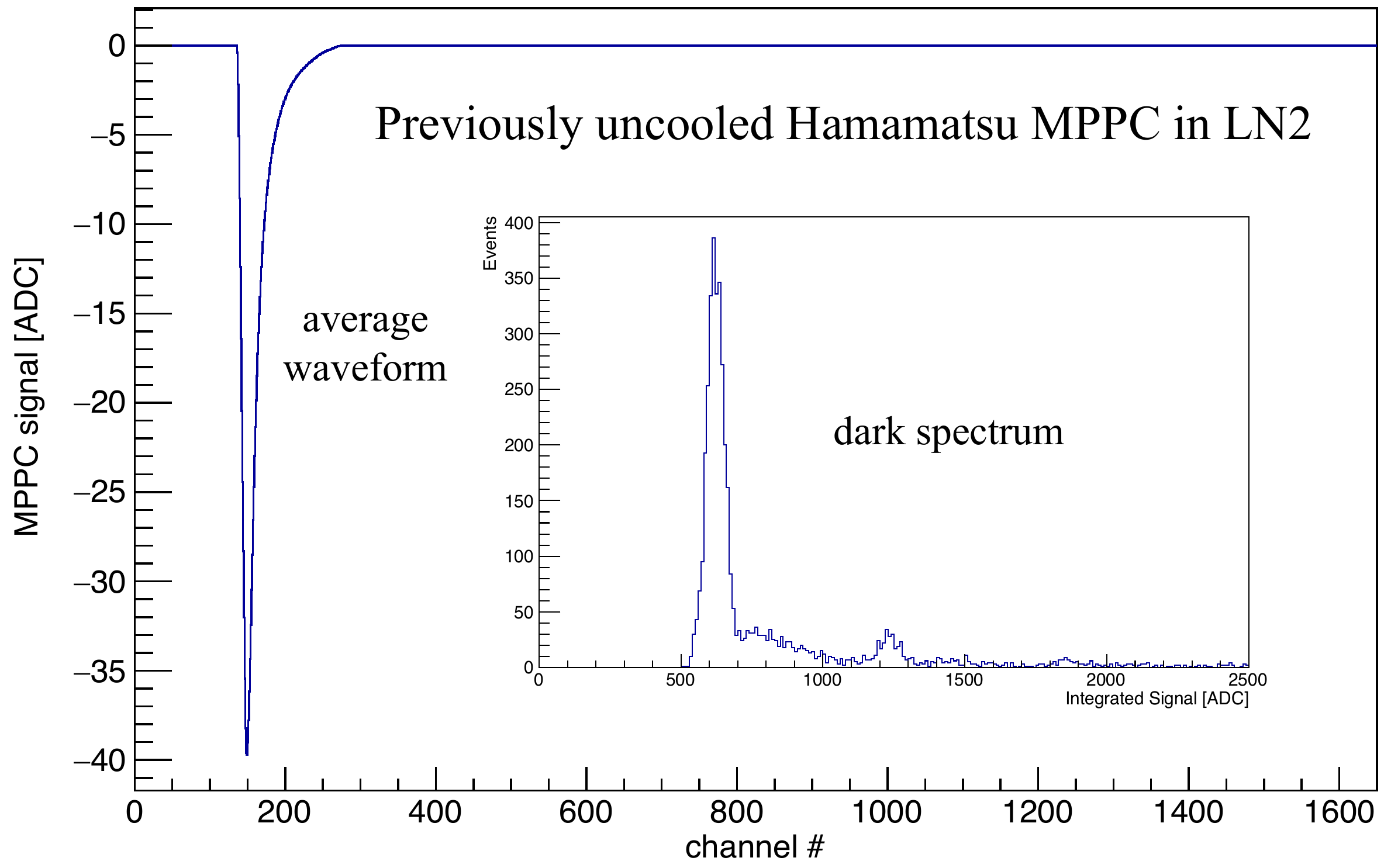}
\includegraphics[width=2.3in,height = 2.1in]{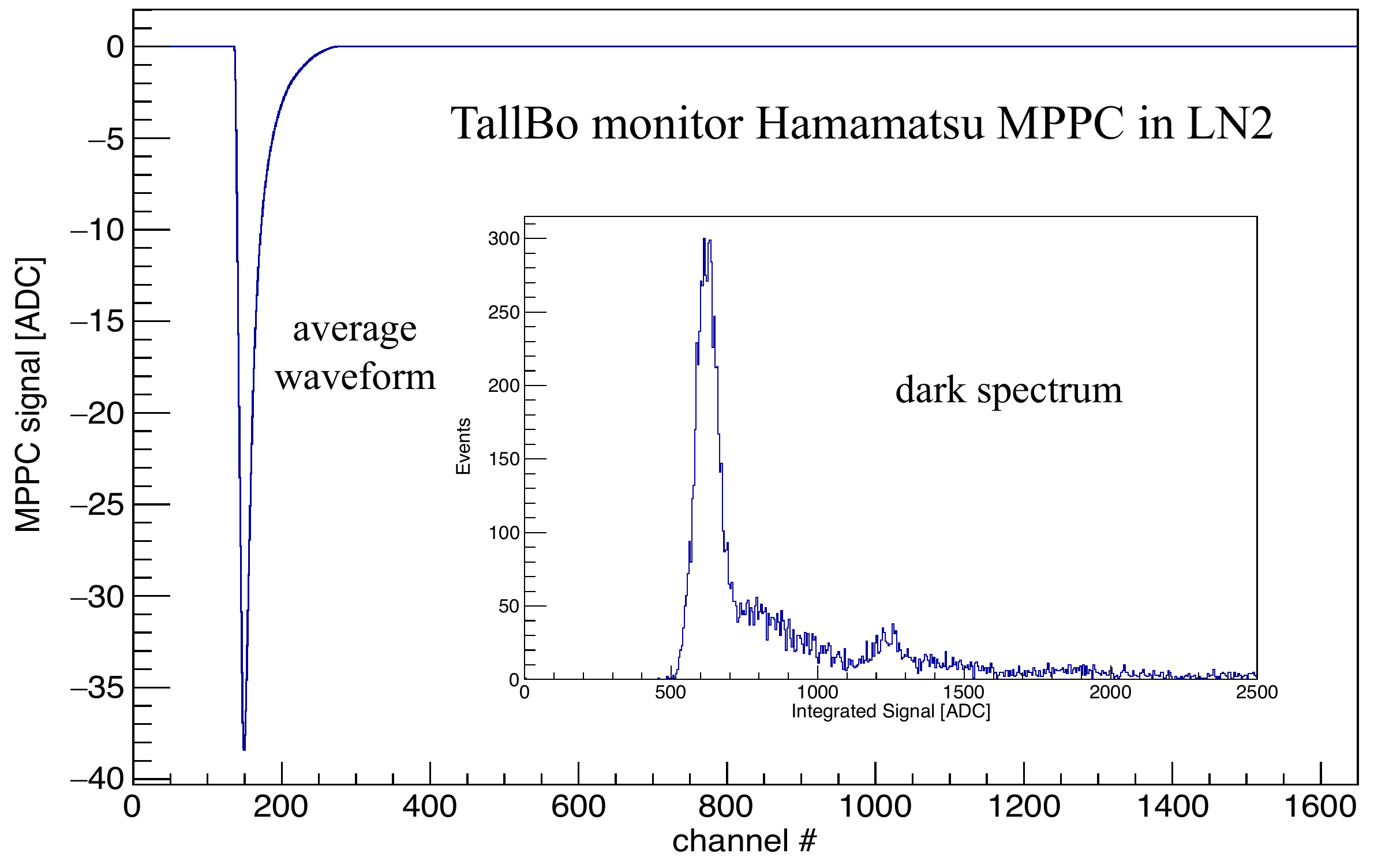}\\
\caption{The comparison of the dark spectrum of a previously uncooled MPPC in LN2 biased at 44.5~V with the dark spectrum of the monitor MPPC also biased at 44.5~V.  Both spectra are superposed on an average of 25 dark waveforms.  These MPPCs demonstrate behavior that is quite similar. }
\label{compMPPCsAppendix}
\end{figure}

Fig.~\ref{compMPPCsAppendix} shows the comparison of the dark spectrum of a previously uncooled MPPC in LN2 biased at 44.5~V (cf., Fig.~\ref{testMPPCAppendix}) compared with the dark spectrum of the monitor MPPC also biased at 44.5~V.  These spectra are superposed on an average of 25 dark waveforms.  There does seem to be some evidence for degradation in the dark spectra of the monitor MPPC, and by extension the MPPCs used on the readout boards.  The first p.e. peak peak is somewhat reduced in the monitor MPPC and the afterpulsing has increased.  The integrated spectra are estimated to differ by less than 10$\%$.  

\newpage

\subsection{Read Out Boards}

This figure shows the baseline subtracted mean of 25 waveforms from single track muon events crossing TallBo as read out by a passive ganging board in runs 2 (front), run 3 (back), and run 4 (front-reversed).   The waveforms are offset from one another for clarity. 
\begin{figure}[h]
\centering
\includegraphics[width=2.3in,height = 2.1in]{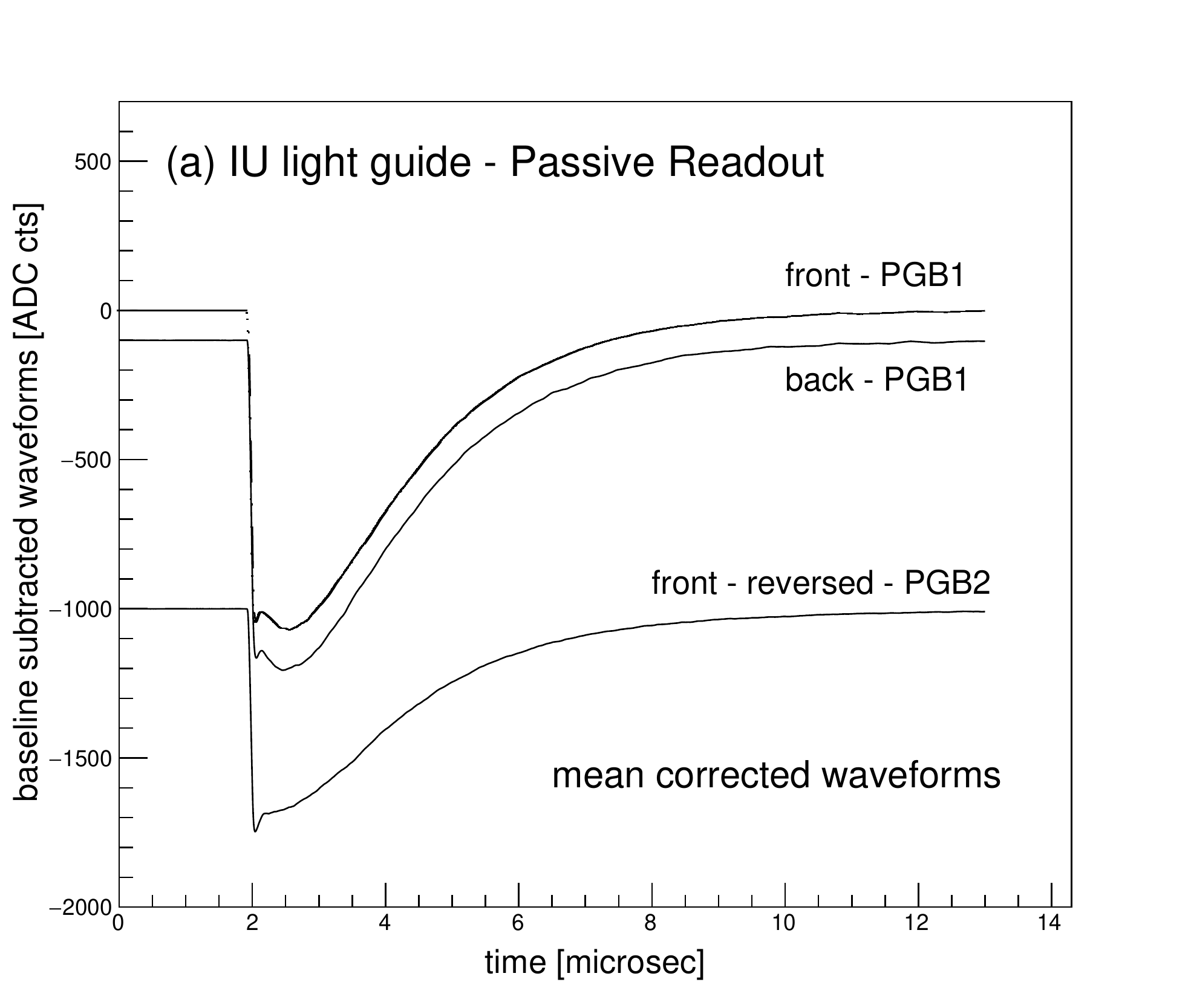}
\includegraphics[width=2.3in,height = 2.1in]{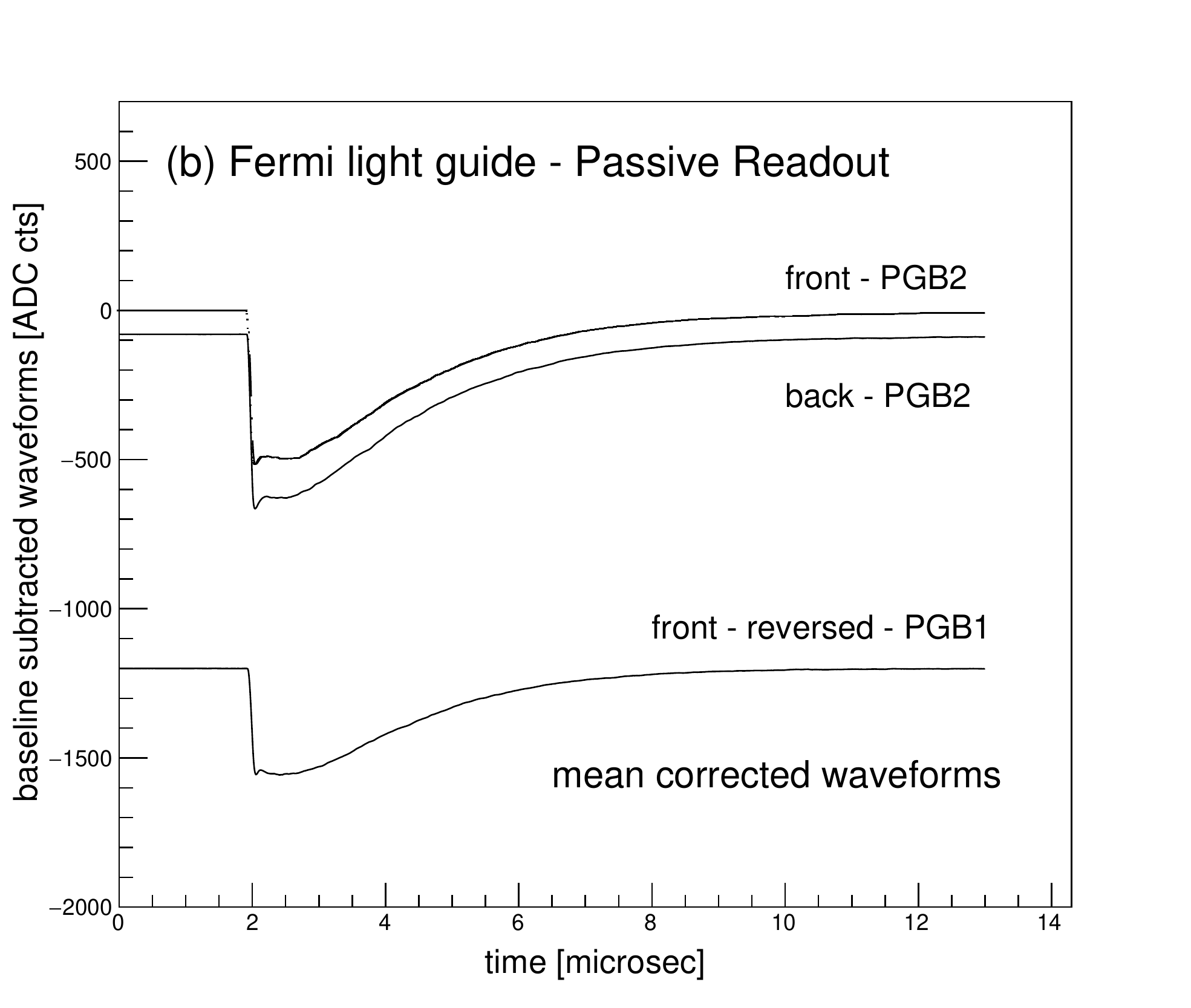}\\
\caption{The baseline subtracted mean of 25 waveforms from single track muon events crossing TallBo as read out by a passive ganging board in runs 2 (front), run 3 (back), and run 4 (front-reversed).  The waveforms are offset from one another for clarity and are labeled with the readout boards that collected them.  The waveforms in the figure have been multiplied by the geometry correction given in Table~\ref{tab:Results}.}
\label{passiveBoardAppendix}
\end{figure}

\begin{figure}[h]
\centering
\includegraphics[width=2.3in,height = 2.1in]{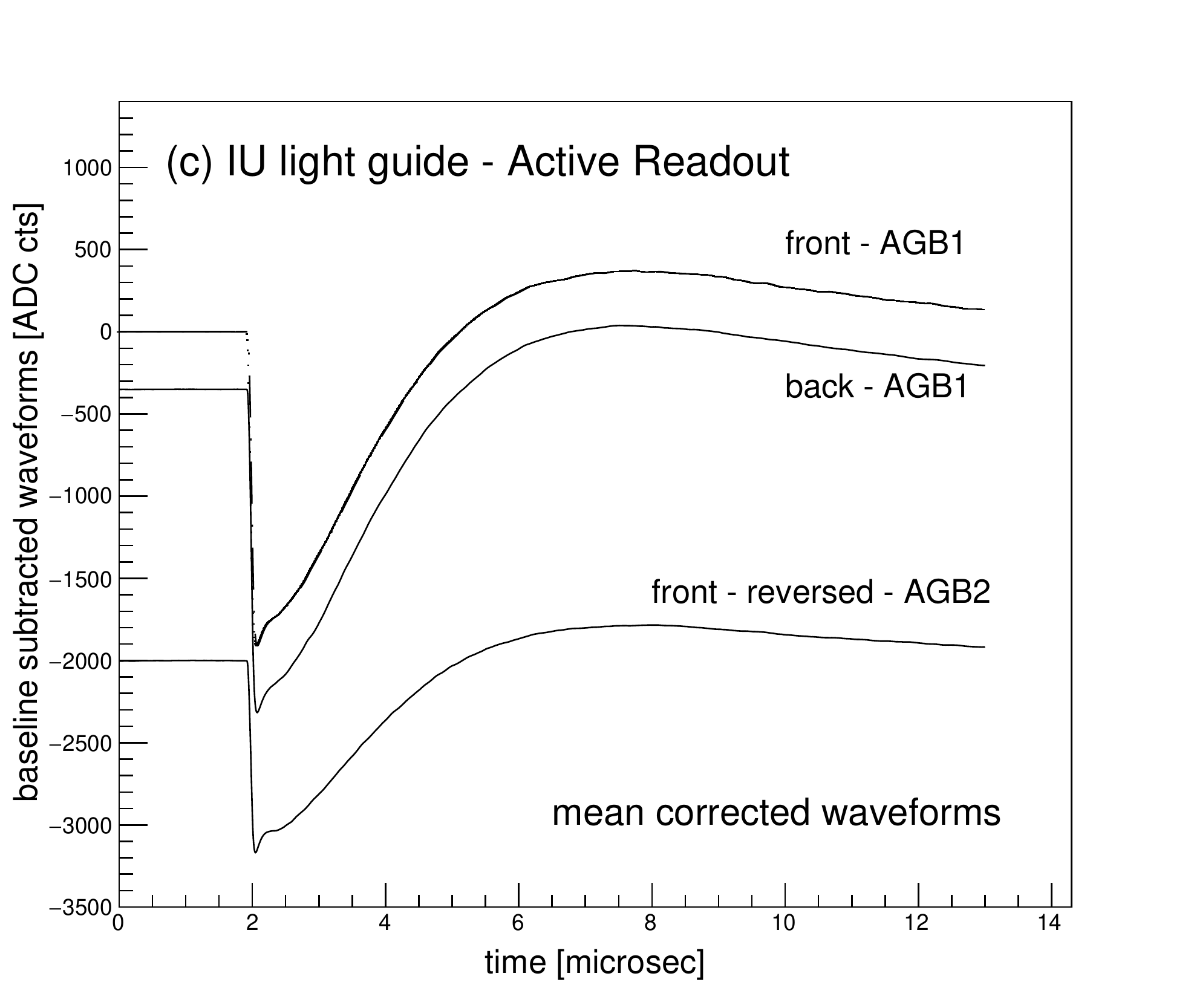}
\includegraphics[width=2.3in,height = 2.1in]{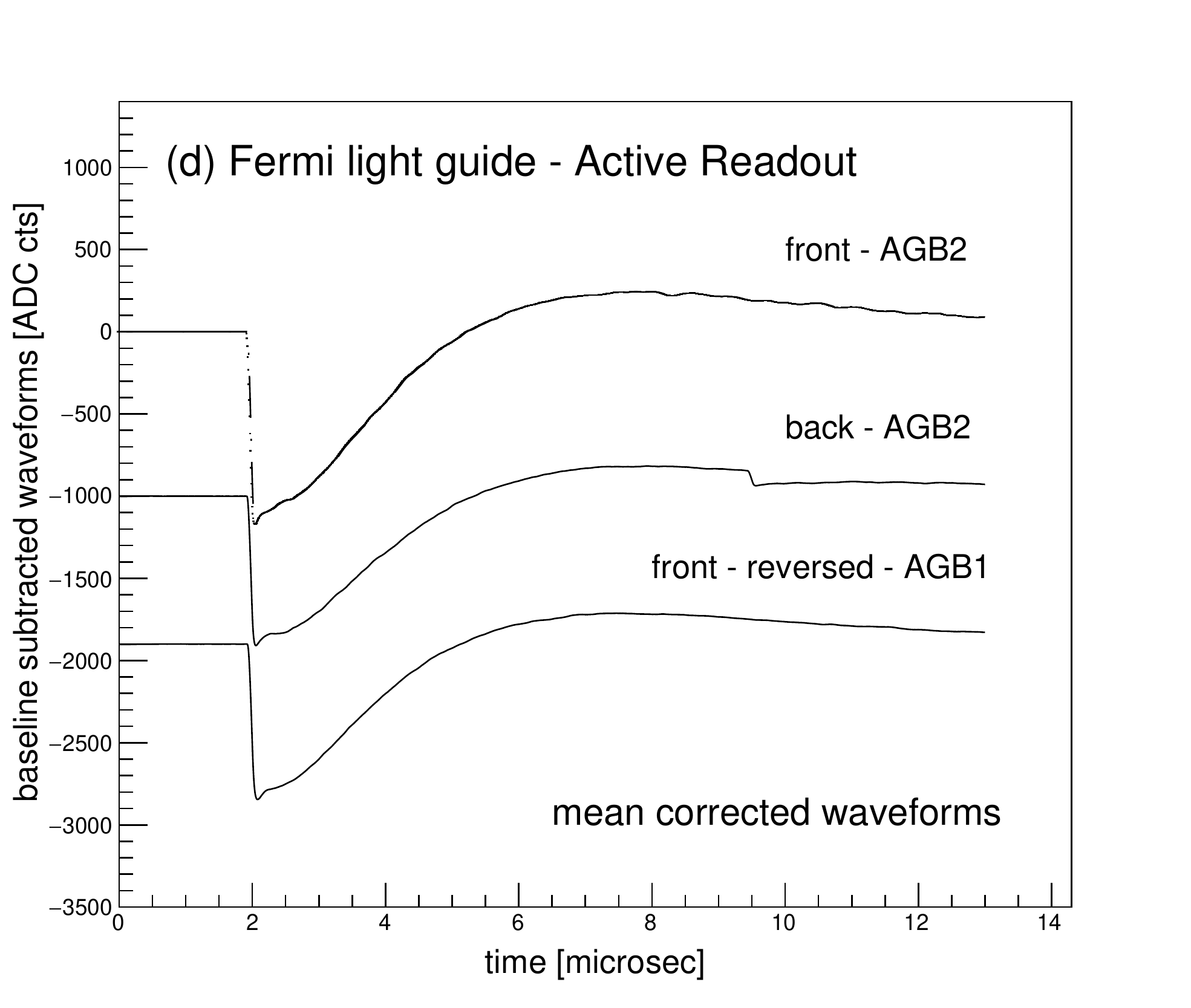}\\
\caption{The baseline subtracted mean of 25 waveforms from single track muon events crossing TallBo as read out by an active ganging board in runs 2 (front), run 3 (back), and run 4 (front-reversed).  The waveforms are offset from one another for clarity and are labeled with the readout boards that collected them.  The waveforms were multiplied by the geometry correction given in Table~\ref{tab:Results}.}
\label{activeBoardAppendix}
\end{figure}

Fig~\ref{activeBoardAppendix} shows the baseline subtracted mean of 25 waveforms from single track muon events crossing TallBo as read out by an active ganging board in runs 2 (front), run 3 (back), and run 4 (front-reversed).  


\subsection{Difference in Track Length}
\label{differenceInTrackLengthAppendix}

The difference in the [ADC]/track from run 2 to run 4 could result from a difference the number of photons striking the light guides due to different path length distributions in LAr for the two runs.  
 A cosmic track between fixed PMT positions on the hodoscopes can be characterized by its zenith and azimuthal angles and every track with these angles passes through a specific path length of LAr.  Consequently, the zenith angle and azimuthal angle distributions for the tracks correlate with the path length distributions of the cosmics in LAr and the scintillation photons given off.  The zenith angle and azimuthal angle distributions for the tracks passing the cuts from the passive readout boards in runs 2 and 4 are shown in Fig.~\ref{Distributions}.
As can be seen, these distributions are quite similar.  The track length distributions are also shown in Fig.~\ref{Distributions}.  As for the angle distributions, the track length distributions are quite similar.  The differences in these distributions can be attributed mostly to the stochastic nature of cosmic rays and the different statistics in the two runs.  Fig.~\ref{Distributions} suggests that the drop-off in signal seen from run 2 to run 4 are unlikely to be from differences in the path length distributions between the two runs.
\begin{figure}[h!]
\centering
\includegraphics[width=2.3in,height = 1.9in]{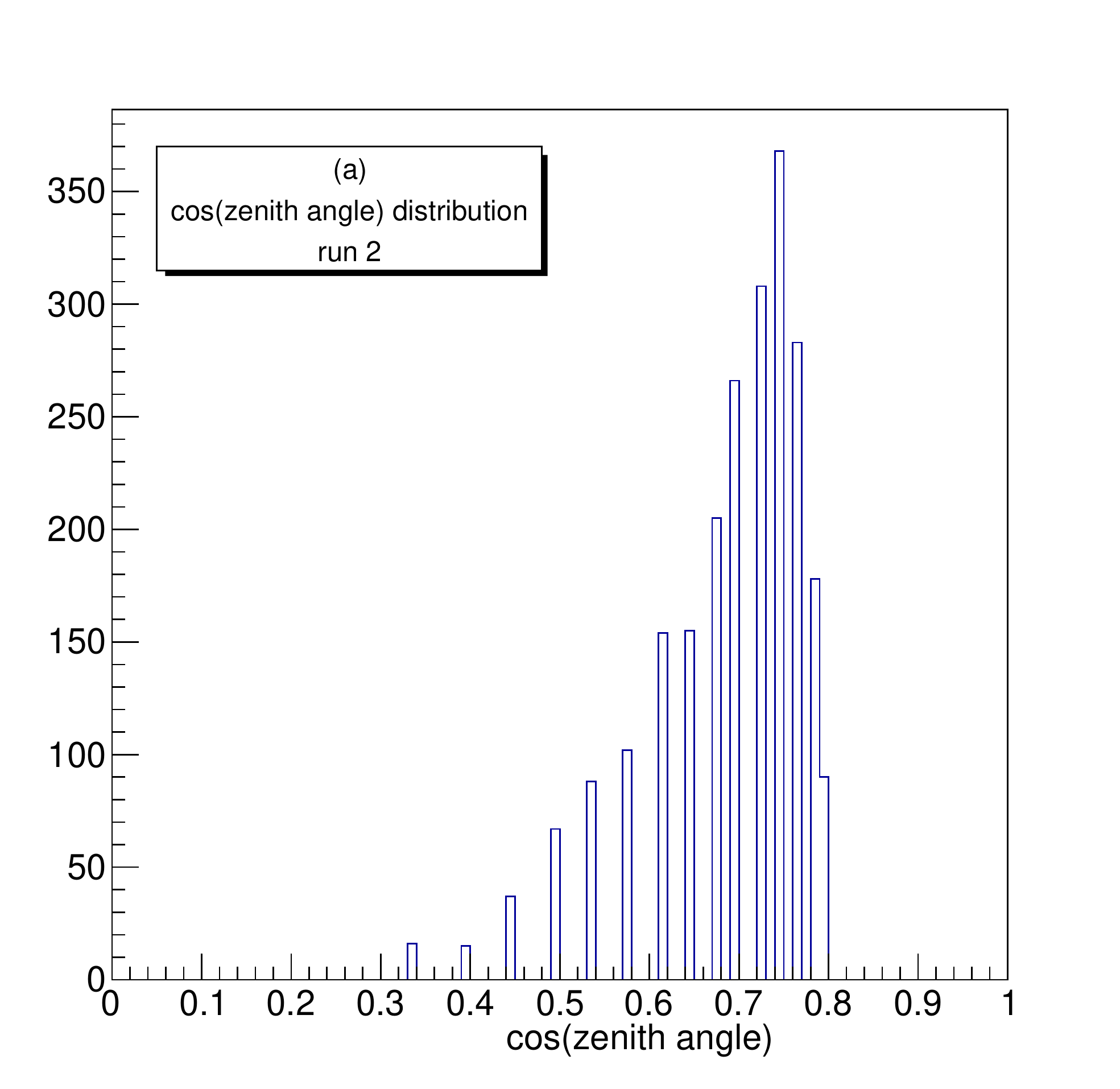}
\includegraphics[width=2.3in,height = 1.9in]{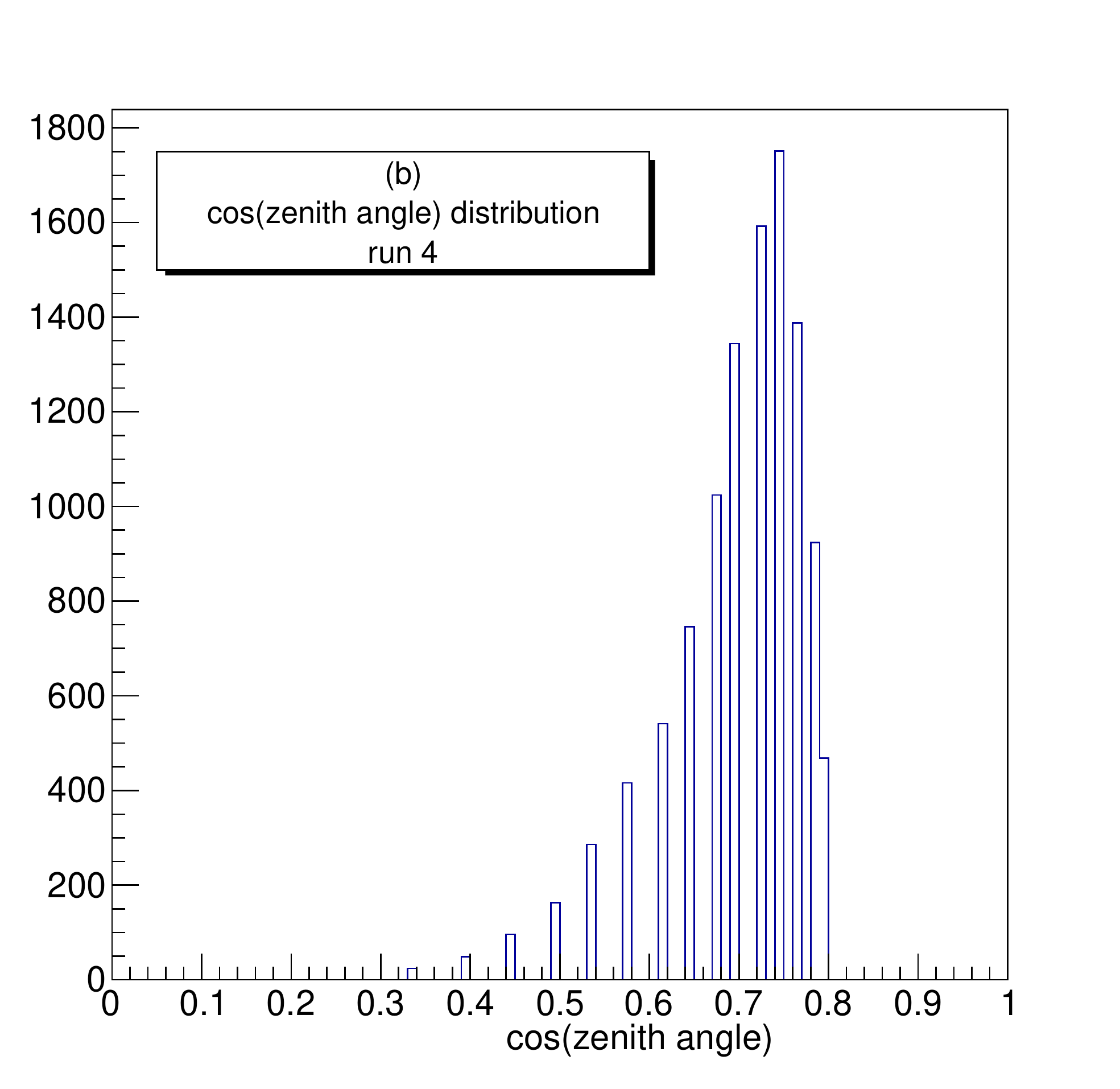}\\
\includegraphics[width=2.3in,height = 1.9in]{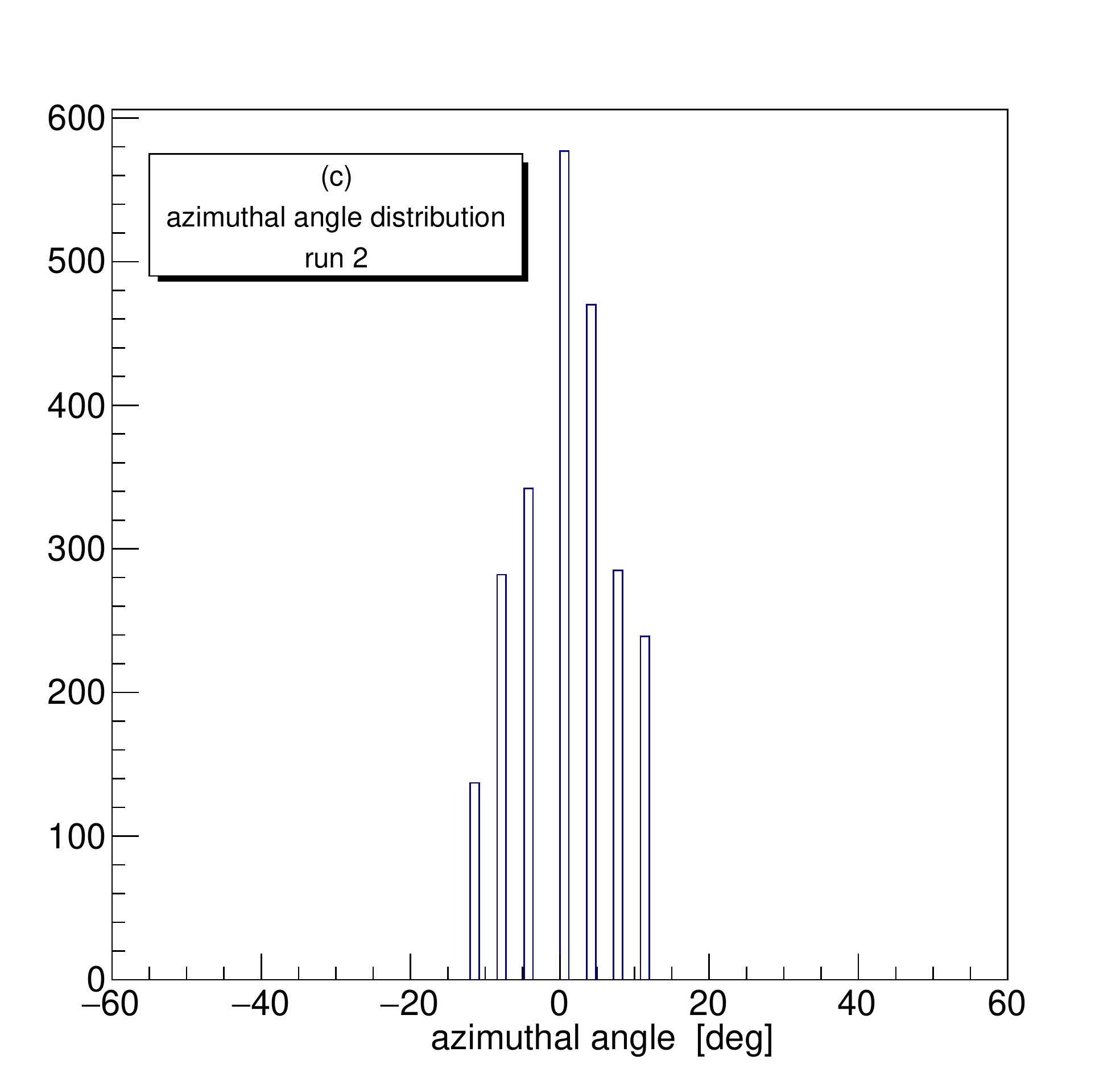}
\includegraphics[width=2.3in,height = 1.9in]{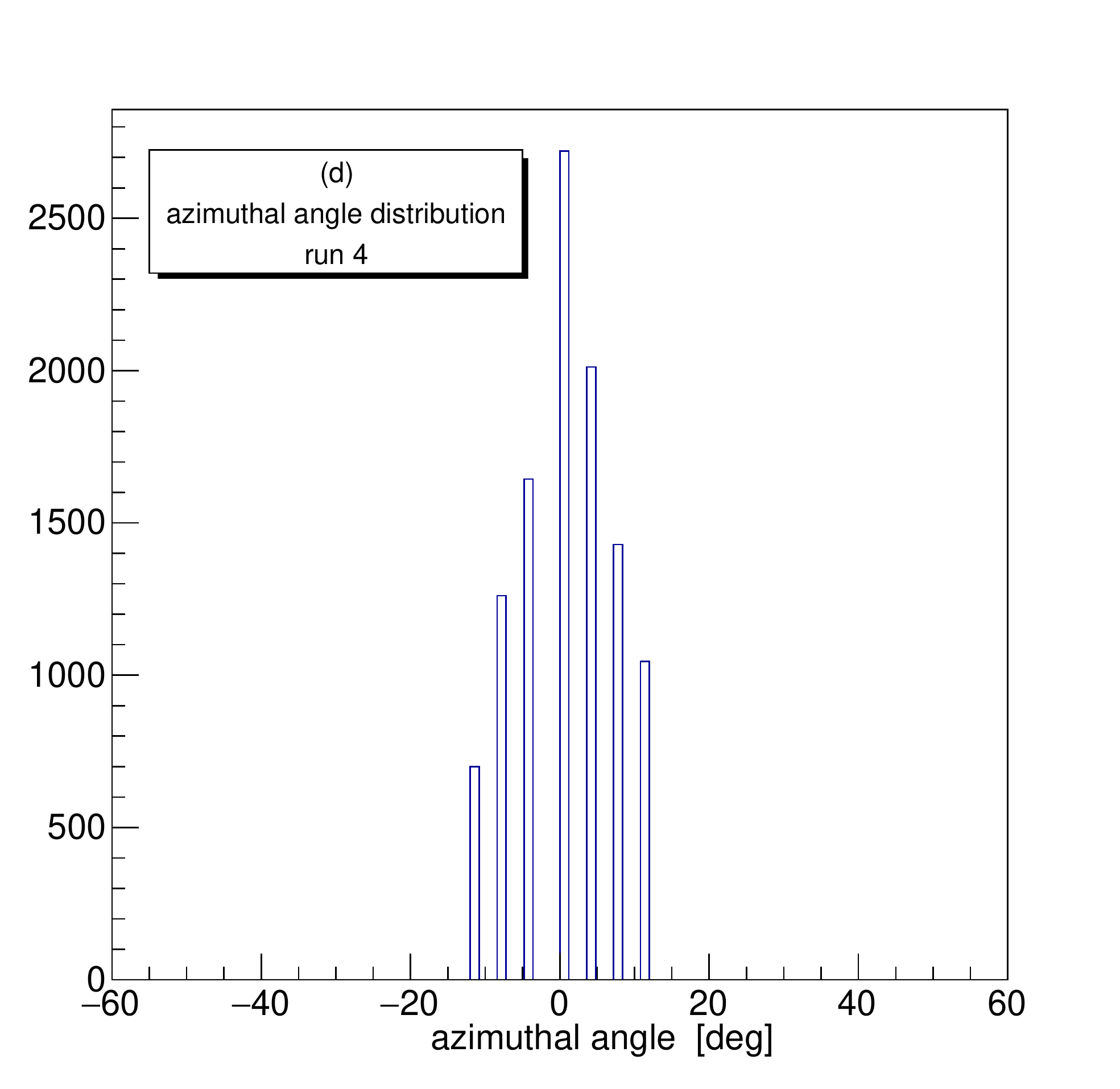}\\
\caption{(a) the zenith angle distribution in run 2; (b) the zenith angle distribution in run 4; (c) the azimuthal angle distribution in run 2; (d) the azimuthal angle distribution in run 4.}
\label{DistributionsAppendix}
\end{figure}

\end{document}